\documentclass[submission, Phys]{SciPost}
\usepackage{graphicx,amsmath,amsfonts,amssymb,amsthm,xr}

\usepackage{epsfig,color,dsfont,upgreek,physics}
\usepackage{blkarray,bigstrut}
\usepackage{mathrsfs}
\usepackage{mathtools}
\usepackage{bbold}

\usepackage{float}

\usepackage{hyperref}
\usepackage{xcolor}
\usepackage{orcidlink}

\usepackage[normalem]{ulem}

\newcommand{\sfigref}[2]{Fig.\,\hyperref[#1]{\ref{#1}#2}}

\definecolor{kspink}{RGB}{200,0,200}

\hypersetup{
  colorlinks = true,
  urlcolor = blue,
  pdfauthor = {Sayak Guha Roy Kevin Slagle},
  pdftitle = {Guha Roy and Slagle - 2026 - Reweighted Time-Evolving Block Decimation for Improved Quantum Dynamics Simulations}
}

\begin{document}

\begin{center}{\Large \textbf{
Reweighted Time-Evolving Block Decimation for Improved Quantum Dynamics Simulations
}}\end{center}

\begin{center}
\textbf{Sayak Guha Roy\textsuperscript{1,3,$\star$},
Kevin Slagle\textsuperscript{1,2}$^\dagger$}
\end{center}

\begin{center}
{\bf 1} Department of Physics and Astronomy, Rice University, Houston, Texas 77005, USA
\\
{\bf 2} Department of Electrical and Computer Engineering, Rice University, Houston, Texas 77005 USA
\\
{\bf 3} Smalley-Curl Institute, Rice University, Houston, Texas 77005, USA

$\star$ \href{mailto:sg161@rice.edu}{sg161@rice.edu} \\
$\dagger$ \href{mailto:kevin.slagle@rice.edu}{kevin.slagle@rice.edu}
\end{center}

\begin{center}
\today
\end{center}

\section*{Abstract}
{\bf
We introduce a simple yet significant improvement to the time-evolving block decimation (TEBD) tensor network algorithm for simulating the time dynamics of strongly correlated one-dimensional (1D) mixed quantum states. The efficiency of 1D tensor network methods stems from using a product of matrices to express either: the coefficients of a wavefunction, yielding a matrix product state (MPS); or the expectation values of a density matrix, yielding a matrix product density operator (MPDO). To avoid exponential computational costs, TEBD truncates the matrix dimension while simulating the time evolution. However, when truncating an MPDO, TEBD does not favor the likely more important low-weight expectation values, such as $\langle c_i^\dagger c_j \rangle$, over the exponentially many high-weight expectation values, such as $\langle c_{i_1}^\dagger c^\dagger_{i_2} \cdots c_{i_n} \rangle$ of weight $n$, despite the critical importance of the low-weight expectation values. Motivated by this shortcoming, we propose a reweighted TEBD (rTEBD) algorithm that deprioritizes high-weight expectation values by a factor of $\gamma^{-n}$ during the truncation. This modification makes rTEBD significantly more accurate than the TEBD time-dependent simulation of an MPDO, and competitive with and sometimes better than TEBD using MPS. Furthermore, by prioritizing low-weight expectation values, rTEBD preserves conserved quantities to high precision.
}

{
\hypersetup{linkcolor=black}
\vspace{10pt}
\noindent\rule{\textwidth}{1pt}
\tableofcontents\thispagestyle{fancy}
\noindent\rule{\textwidth}{1pt}
\vspace{10pt}
}

\section{Introduction}
\label{sec:intro}

Classical simulation of one-dimensional quantum systems has been a well-studied area over the years with the advent of several simulation techniques. Simulation of matrix product states (MPS) using the time-evolving block decimation (TEBD) \cite{Vidal1,MPS_brickwork} algorithm is one such technique. An advantage of using an MPS over a full Schr\"odinger-picture state vector is the ability to efficiently encode the wavefunction amplitudes using matrix products. Ref.~\cite{Vidal1} showed that the MPS is an accurate approximation for quantum systems with low entanglement.
However, time-evolving quantum systems are still a major challenge because the quantum entanglement grows linearly with time, making the MPS representation inaccurate at large times. To tackle this issue, several new time evolution algorithms have been developed, such as Local-Information Time Evolution \cite{local-info1,local-info2}, time evolution using Density Matrix Truncation (DMT) \cite{White_2018,PhysRevLett.125.030601}, time evolution in the Heisenberg picture using Dissipation Assisted Operator Evolution (DAOE) \cite{DAOE,srivatsa2024}, fermion DAOE for free or weakly interacting fermions \cite{lloyd2023,kuo2023}, sparse Pauli dynamics \cite{sparse}, operator-size truncated (OST) dynamics \cite{PhysRevB.110.134308} and universal operator growth hypothesis \cite{PhysRevX.9.041017}. The growth of entanglement with time in quantum systems mostly comes from higher weight correlations which do not play a major role in the hydrodynamics and many of the recent methods developed rely on somehow neglecting such terms \cite{PhysRevB.105.245101,PhysRevB.107.094311}. Efforts have also been made to reduce the entanglement using purification \cite{Hauschild_Leviatan_Bardarson_Altman_Zaletel_Pollmann_2018}. Additionally, variational time evolution of tensor networks (TDVP methods) \cite{PRXQuantum.5.020361,Bauernfeind_Aichhorn_2019,Goto_Danshita_2019,Haegeman_Cirac_Osborne_Pižorn_Verschelde_Verstraete_2011}, have been developed that focus on approximately conserving the total energy and preserving the wavefunction norm by projecting the dynamics onto the tangent space of the variational manifold. Related tangent-space approaches have been used to compute real-frequency spectral functions \cite{tangent-space-krylov}. Furthermore, there has been significant progress in the use of tensor networks to study time evolution and steady state properties of many-body quantum systems \cite{DMRG-1,Daley_2004,McCulloch_2007,to_mps3,to_mps4,to_mps5,Kshetrimayum_Weimer_Orus_2017,Bañuls_Hastings_Verstraete_Cirac_2009,Czarnik_Dziarmaga_Corboz_2019}, and MPS methods have been used in several theoretical \cite{roy2025repulsivelyboundhadronsmathbbz2,10.21468/SciPostPhys.16.5.138,kondo-app,TDVP-1,URBANEK2016170,Haegeman_Lubich_Oseledets_Vandereycken_Verstraete_2016,Haegeman_Osborne_Verstraete_2013,PhysRevB.100.104303,PhysRevB.86.245107,Zauner_2015} and experimental studies \cite{Mi_2024,exp2}.

Recently, there have been advances in using a matrix product operator (MPO) to encode a time evolved observable or a matrix product density operator (MPDO) to encode a time evolved density operator \cite{White_2018,Wtwo_stepper,MPO-1,mpdo-2,mpdo-3,Cui_Cirac_Bañuls_2015}. MPOs and MPDOs are also represented using a matrix product, and they can also be time evolved using the TEBD time evolution scheme \cite{mpdo-tebd,mpdo-tebd2}. The advantage of using an MPO or MPDO over an MPS for time-evolving quantum states lies in the ability of an MPO or MPDO to more efficiently encode quantum entanglement and correlations.

However, the TEBD time-evolution of an MPDO does a poor job at conserving conserved quantities (including the rather trivial density matrix trace).
Ref. \cite{White_2018} proposes a clever and fairly simple Density Matrix Truncation (DMT) method, which can conserve local conserved quantities nearly exactly (i.e. up to machine precision errors).
TEBD utilizes a singular value decomposition (SVD) truncation as the approximate step to limit computational cost.
In the simplest case, DMT modifies this SVD truncation such that all nearest-neighbor 3-body expectation values are not changed by the truncation approximation, which allows conserved quantities that are a sum of nearest-neighbor 3-body operators to be exactly conserved. 
This allows DMT to more accurately capture long time hydrodynamic and complex intermediate time behavior.
However, the truncation approximation used by DMT fails to maintain longer-range two-body correlations, such as $\langle\sigma_i^\mu \sigma_{i+3}^\nu\rangle$.
This occurs because DMT and TEBD give simple two-body correlations such as $\langle\sigma_i^\mu \sigma_{i+3}^\nu\rangle$
exactly the same priority in the approximation step as many-point correlation functions, such as $\langle\sigma_1^{\mu_1} \cdots \sigma_{n}^{\mu_n}\rangle$.
Since there are exponentially many many-point correlation functions, the 2-point correlations are quickly forgotten by the MPDO.

To address this shortcoming, we develop a new time evolution technique, called the Reweighted Time Evolving Block Decimation (rTEBD). The algorithm is similar to a TEBD time evolution of a MPDO. However, in our algorithm, we reweight the MPDOs using a reweighted Pauli basis such that correlation functions involving $n$ Pauli operators are reweighted, or prioritized, by a factor of $\gamma^{-n}<1$ during the SVD truncation.
This reweighting allows rTEBD to maintain few-point correlations significantly better,
  allowing for more accurate quantum dynamics simulations.
Conserved quantities, such as the total energy, are also approximately conserved as a result.

The DAOE method \cite{DAOE,srivatsa2024} and the Complex time-evolution schemes \cite{complex-time,complex-time2,complex-time3} were also developed along a similar motivation of preserving few-point correlations better. DAOE introduces a dissipation superoperator that exponentially suppresses higher-weight Pauli strings while complex time-evolution schemes replace the time-evolution unitary with $\exp[{-(i+\epsilon)(\mathcal{H}-E_0)\delta t}]$, introducing a controlled non-unitary component that suppresses high-energy contributions and can reduce entanglement growth in numerical simulations. Recent work has further developed Krylov-based variants of this idea to extend the accessible time domain in spectral function calculations \cite{complex-time-krylov}. Since rTEBD modifies only the truncation step of standard TEBD, its implementation is more straightforward than DAOE. Furthermore, the time evolution in DAOE and the complex time-evolution schemes is non-unitary and typically requires post-processing such as unitary extrapolation to recover physical observables. In contrast, rTEBD leaves the underlying time evolution unitary (up to Trotter and SVD errors) and biases only the SVD truncation through the Pauli-basis reweighting. This makes rTEBD conceptually distinct from complex-time methods, and these approaches operate at different stages of the simulation and could in principle be combined. A different strategy with similar underlying motivation is the Local Information Time Evolution (LITE) method, which divides the system into subsystems and exactly time-evolves them, discarding information beyond a chosen subsystem size. A systematic comparison with these methods is an interesting direction for future work.

Another complementary direction for time evolution of MPS is using global time evolution Matrix Product Operators (MPOs) constructed directly from the Hamiltonian. The $W^I$ and $W^{II}$ methods \cite{Wtwo_stepper} provide such a construction that naturally accommodates long-range Hamiltonians without gate swaps or Trotter restructuring. These MPO-based steppers are widely used and are particularly advantageous for Hamiltonians beyond nearest-neighbor form. We emphasize that rTEBD modifies the truncation step rather than the time-stepping scheme itself. The Pauli-basis reweighting is therefore distinct from the choice of time stepper: the same reweighting strategy could in principle be combined with MPO-based steppers such as $W^{II}$ \cite{Wtwo_stepper}, by replacing the standard time-evolution MPO with its reweighted super-operator analogue. A systematic comparison and integration of the reweighting scheme with these approaches is an interesting direction that we leave to future work.

\begin{figure}
    \centering
    \includegraphics[width=0.7\linewidth]{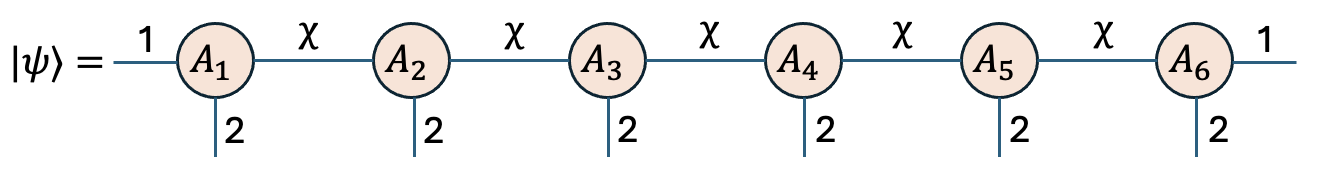}
    \caption{Representation of a wavefunction of a 6-site qubit chain using a matrix product state.}
    \label{fig:MPS_1}
\end{figure}

We outline our paper in the following way. In section \ref{sec:review}, we review the MPS, MPDO, and the TEBD algorithms.
In  section \ref{sec:rewt}, we introduce the new rTEBD algorithm for bosonic and fermionic systems. In section \ref{sec:benchmark}, we benchmark rTEBD against TEBD of MPS and MPDO for a free fermion system (for which exact solutions are known but which are no harder than interacting systems for these algorithms). rTEBD performs significantly better than MPDO-TEBD and sometimes slightly better than MPS-TEBD. We conclude in Section.~\ref{sec:conc}.

\section{Review}
\label{sec:review}

Before explaining the rTEBD algorithm, we review the TEBD time evolution of MPS and MPDO.

\subsection{MPS}

\begin{figure}
    \centering
    \includegraphics[width=0.7\linewidth]{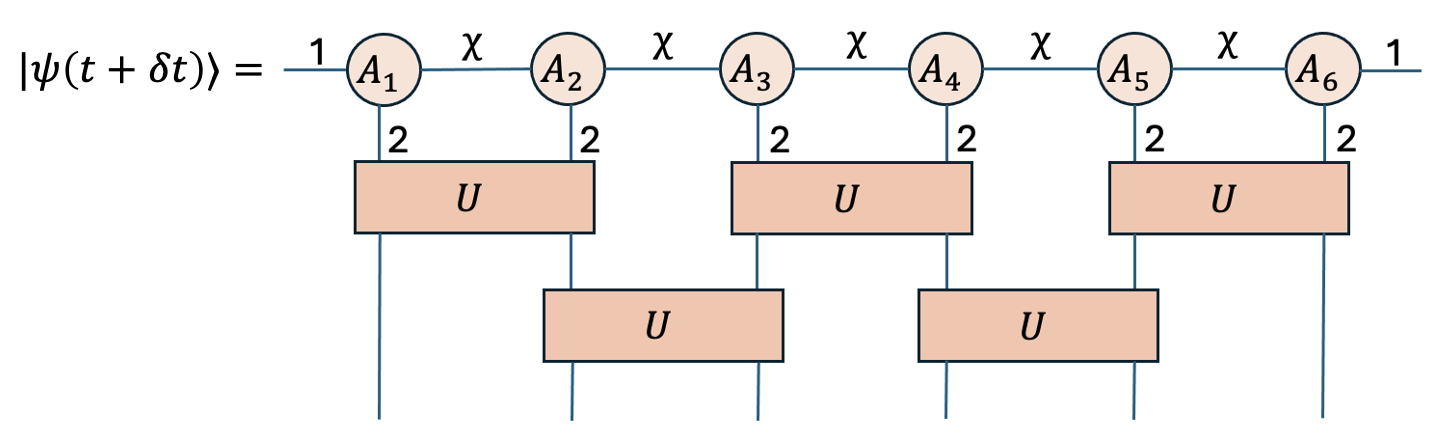}
    \caption{Time evolution of a MPS using TEBD.
    $U$ are the two-qubit unitaries that act in a Trotter decomposed \cite{Childs_2021}  brickwork-like circuit.
    }
    \label{fig:MPS_evolve}
\end{figure}

For a qubit chain, a wavefunction can be written in terms of a matrix product state as shown in Fig. \ref{fig:MPS_1}.
For an $L$-site system, a MPS encodes the wavefunction as the following matrix product of tensors:
\begin{equation}
    \ket{\psi} = \sum_{s_1,s_2,\cdots,s_L} \Tr (A_1^{s_1}A_2^{s_2}\cdots A_L^{s_L})\ket{s_1s_2 \cdots s_L}
\end{equation}
Here, $s_i=\uparrow,\downarrow$ indexes the different spin states.
$A_i^\uparrow$ and $A_i^\downarrow$ are $\chi_i\times\chi_{i+1}$ matrices.
$\chi_i$ is an integer bond dimension.

For a $L$-site qubit system, a bond dimension $\chi=2^{L/2}$ suffices to exactly describe any wavefunction (corresponding to maximal entanglement across the central bipartition). 
An SVD can be used to truncate the bond dimension to approximate the wavefunction. Hence, instead of storing $2^L$ numbers for an $L$-site qubit chain, an MPS only needs to store $Ld \chi^2$ numbers (where $d$ is the local Hilbert space dimension). One can do such an approximation because, for systems following an area-law of entanglement, the relevant physics is captured by states living in a polynomially-small subset of Hilbert space \cite{Hastings_2007}.

\subsection{TEBD of an MPS}
\label{sec:MPS-TEBD}

The time-evolving block decimation (TEBD) algorithm to time evolve a MPS applies 2-qubit unitaries based on the system Hamiltonian in a Trotterized \cite{Childs_2021} quantum circuit, followed by SVD to truncate the bond dimension. An example of such a Trotter decomposed time evolution is shown in Fig. \ref{fig:MPS_evolve}. The application of the 2-qubit unitaries in a brickwork like circuit shown in Fig. \ref{fig:MPS_evolve} simulates the time evolution of the wavefunction and returns the wavefunction at time $t+\Delta t$, where $\Delta t$ is the Trotter step. The brickwork (odd/even) Trotterized circuit is frequently used in the literature for Matrix Product State simulations \cite{Vidal1,MPS_brickwork,brickwork1,brickwork2}.
After each two-qubit unitary, a singular value decomposition (SVD) 
and truncation is used to keep the bond dimension from growing.

\subsection{MPDO}

\begin{figure}
    \centering
    \includegraphics[width=0.7\linewidth]{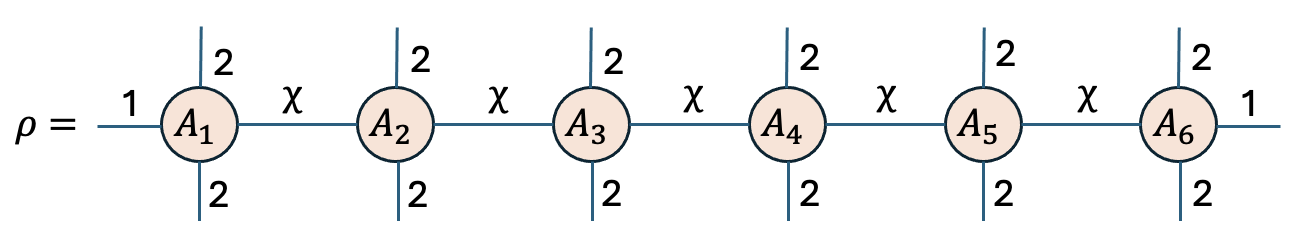}
    \caption{Representation of a density matrix $\rho$ in terms of a matrix product density operator (MPDO).}
    \label{fig:MPDO_1}
\end{figure}

Similar to MPS, where we write a wavefunction as a matrix product, we can also write a density matrix as a matrix product. This is known as a matrix product density operator (MPDO), depicted in Fig. \ref{fig:MPDO_1}.

For us, it is useful to express the MPDO in a Pauli basis (Fig \ref{fig:mpdo_pauli}):
\begin{equation}
    \rho(t) = \frac{1}{2^L}\sum_{\mu_1,\cdots ,\mu_L} \sigma^{\mu_1}\otimes \sigma^{\mu_2} \otimes \cdots  \otimes \sigma^{\mu_L} A_1^{\mu_1}A_2^{\mu_2}\cdots A_L^{\mu_L}
    \label{eq:MPDO_pauli}
\end{equation}
where each $A_i^{\mu_i}(t)$ is a function of time. The Pauli basis includes $\sigma^0 \equiv I$, $\sigma^x$, $\sigma^y$, and $\sigma^z$ matrices.
We suppress the matrix trace from now on since the matrix product $A_1^{\mu_1}A_2^{\mu_2}\cdots A_L^{\mu_L}$ results in a scalar $1\times1$ matrix.

\subsection{TEBD of an MPDO}
\label{sec:MPDO-TEBD}

\begin{figure}
    \centering
    \includegraphics[width=8cm]{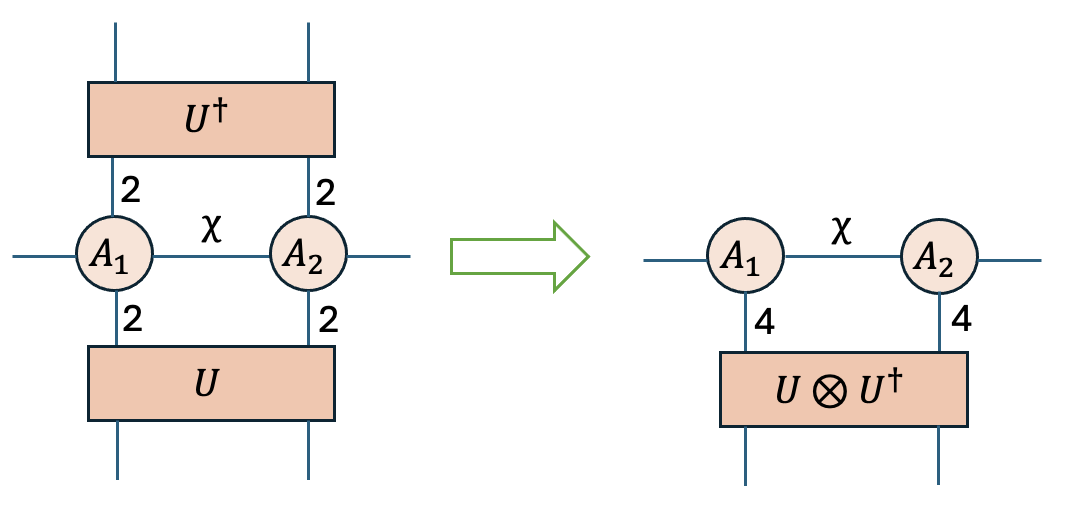}
    \caption{Representation of a MPDO in the Pauli basis.
    The pair of dimension-2 legs of each $A_j$ matrix is combined into a single dimension-4 leg to index the 4 Pauli matrices ($\mathbb{1},\sigma^x,\sigma^y,\sigma^z$). Each pair of two-qubit unitaries, $U$ and $U^\dagger$, are also combined via tensor product, $U \otimes U^\dagger$, to obtain a super-operator that acts on a pair of $A_j$. }
    \label{fig:mpdo_pauli}
\end{figure}

The TEBD time evolution of an MPDO involves applying 2-qubit unitaries in a Trotter decomposed \cite{Childs_2021} circuit on both sides of the MPDO.
For an MPDO written in the Pauli basis, each pair of unitaries can be expressed in the Pauli basis and then combined via tensor product, as depicted in Fig. \ref{fig:mpdo_pauli}, such that the MPS-TEBD algorithm can be directly applied to MPDO.

\section{Reweighted TEBD}
\label{sec:rewt}
In this section, we detail the rTEBD algorithm. To clarify nomenclature, throughout this article, we use ``weight" in the standard sense of Pauli operators: the number of non-identity Pauli operators in a Pauli string $\sigma^{\mu_1}\otimes \sigma^{\mu_2} \otimes \cdots  \otimes \sigma^{\mu_L}$; a low-weight operator has $n \ll L$ while a high-weight operator has $n$ comparable to $L$. By ``reweighting", we mean rescaling each non-identity Pauli operator by $\gamma \geq 1$, so that a Pauli string of weight $n$ acquires an overall factor of $\gamma^{n}$ in the basis.
\subsection{Motivation}

To elaborate on the motivation behind the rTEBD algorithm, we consider the expansion of a density operator in the Pauli basis:

\begin{equation}
\begin{aligned}
    \rho =& a \mathbb{1} + \sum_{\substack{i \\ \mu}}b_i^{\mu} \sigma_i^{\mu} + \sum_{\substack{i<j \\ \mu,\nu}}c_{ij}^{\mu\nu}\sigma_i^{\mu}\sigma_j^{\nu} \\ &+ \sum_{\substack{i<j<k \\ \mu,\nu,\eta}}d_{ijk}^{\mu\nu\eta}\sigma_i^{\mu}\sigma_j^{\nu}\sigma_k^{\eta} + \cdots
    \label{eq:rho_exp}
\end{aligned}
\end{equation}
where here, $\mu,\nu$, and $\eta$ sum over $x$, $y$, and $z$.
Each term encodes an $n$-point expectation value, such as the 1-point $\langle \sigma_i^\mu \rangle$ or 2-point $\langle \sigma_i^\mu \sigma_j^\nu \rangle$ expectation values.

In many physically relevant regimes, especially hydrodynamics and transport, the long-time dynamics is governed by a small set of lower-weight observables \cite{Hydrodynamics1,Hydrodynamics2}, despite the underlying exponential complexity. In particular, in the hydrodynamic regime, the relevant physics is captured by few-point correlation functions of conserved densities and currents, which encode the slow modes of the system. While unitary dynamics generates an exponentially large set of higher-weight operators, these predominantly encode scrambling and entanglement growth \cite{OTOCs}, rather than directly entering hydrodynamic description. Although modern experimental platforms like quantum gas microscope \cite{QGM1} can access higher-order correlations, standard probes of transport and thermalization are typically governed by lower-weight observables. We note that this prioritization of lower-weight observables may be less appropriate for systems where higher-weight operators are physically essential, such as topologically ordered phases probed by string-order parameters \cite{topology_strings}. 

The SVD truncation for TEBD of an MPDO gives each term (e.g. $\langle\sigma_2^z\rangle$ or $\langle\sigma_2^z\sigma_4^y\sigma_5^x\sigma_6^x\sigma_8^z\rangle$; lower or higher-weight) exactly the same importance, despite the fact that for probing hydrodynamics and transport, the few-point terms are arguably more important \cite{Hydrodynamics1}. This is because, SVD-based truncation chooses the best low-rank approximation in the Frobenius (Hilbert–Schmidt) norm by minimizing $\norm{\rho-\rho_{\chi}}_F$, where $\rho_{\chi}$ is the rank-$\chi$ truncated MPDO returned by the algorithm at a given bipartition. This is the optimal rank-$\chi$ approximation in the Frobenius norm for that bipartition.

The rTEBD method that we develop gives preference to the more hydrodynamically relevant low-weight expectation values so that the SVD truncation more accurately preserves the low-weight expectation values. The idea behind rTEBD is to define the operators and MPDO in a reweighted Pauli basis:
\begin{equation}
    \rho(t) = \frac{1}{2^L}\sum_{\mu_1,\cdots,\mu_L} \Tilde{\sigma}^{\mu_1\cdots\mu_L} A_1^{\mu_1}A_2^{\mu_2}\cdots A_L^{\mu_L}
    \label{eq:mpdo_rewt}
\end{equation}
$L$ is the number of qubits, and $A_1^{\mu_1}A_2^{\mu_2}\cdots A_L^{\mu_L}$ is the matrix product resulting in a complex number. We define
\begin{equation}
    \Tilde{\sigma}^{\mu_1\cdots \mu_L} = \Tilde{\sigma}^{\mu_1}\otimes \Tilde{\sigma}^{\mu_2} \otimes \cdots  \otimes \Tilde{\sigma}^{\mu_L}
\end{equation}
as a tensor product of reweighted Pauli matrices. Reweighting the operators can be thought of as defining a modified inner product $\langle A,B\rangle_R = \Tr{A^{\dagger}R^{\dagger}RB}$ where $R$ is a diagonal reweighting operator in the Pauli basis that scales low-weight sectors differently than high-weight ones. In this reweighted norm, rTEBD’s SVD minimizes $\norm{R^{\dagger}(\rho - \rho_{\chi})R}_F$. This is not equivalent to minimizing the physical Frobenius norm, but is a controlled bias that preserves hydrodynamically relevant low-weight contributions at the truncation step.

Because the physical norm is modified, rTEBD does not minimize global error in the unweighted sense; high-weight operator components may be underrepresented in the reweighted MPDO. This is acceptable in many physical contexts where low-weight operators dominate observables of interest. Hence, the motivation behind rTEBD is that, for physical problems that care about low-weight expectation values, rTEBD is expected to more accurately preserve them.

\subsection{Bosonic/spin systems}
In this section, we introduce the rTEBD algorithm for bosonic/spin systems. The reweighted MPDO is defined according to Eq. \eqref{eq:mpdo_rewt} and the reweighted Pauli matrices for bosonic systems are:
\begin{equation}
    \Tilde{\sigma}^{\mu} = \begin{cases}
        \sigma^0 & \text{if } \mu=0 \\
        \gamma \sigma^{\mu} & \text{if } \mu \neq 0
    \end{cases}
    \label{eq:boson_rewt}
\end{equation}
$\sigma^{\mu}$ are the usual Pauli matrices and $\gamma\geq1$ is the reweighting parameter. 

\begin{figure}
    \centering
    \includegraphics[width=0.7\linewidth]{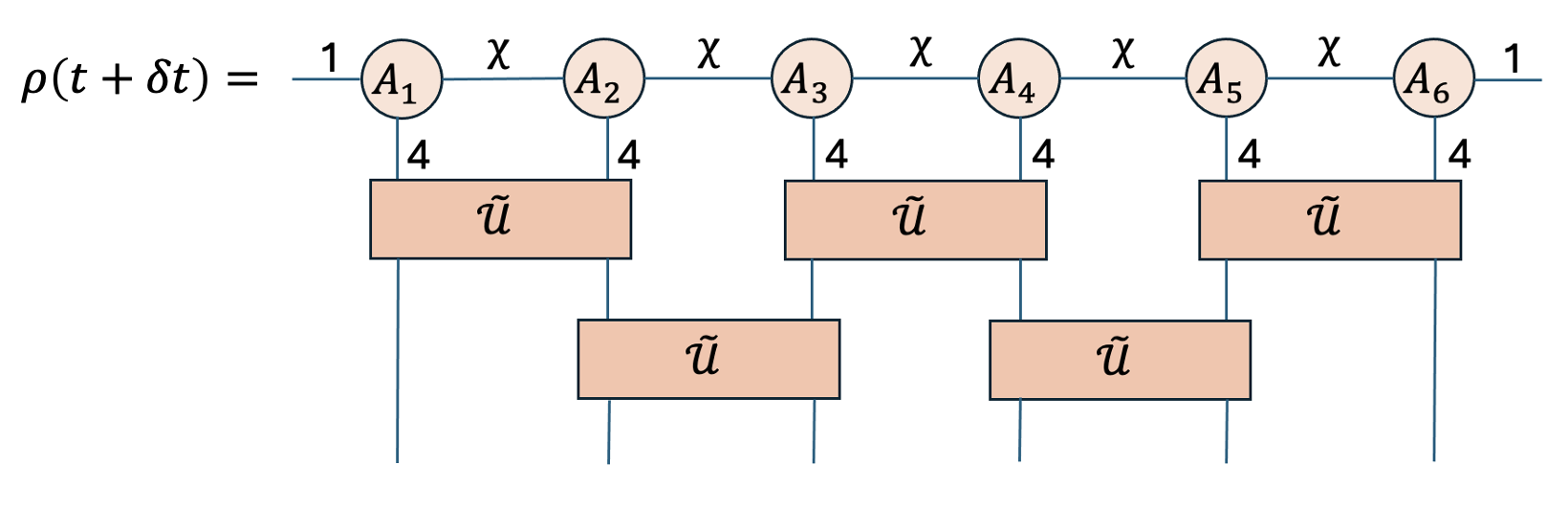}
    \caption{Time evolution of a reweighted MPDO defined in Eq. \eqref{eq:mpdo_rewt} using rTEBD.
    $\Tilde{\mathcal{U}}$ are the two-qubit super-operators defined in the reweighted Pauli basis (Eq. \eqref{eq:boson_rewt}) that act in a Trotter decomposed \cite{Childs_2021}  brickwork-like circuit.
    }
    \label{fig:MPDO_evolve}
\end{figure}

The time evolution of the reweighted MPDOs using TEBD is the rTEBD algorithm. The circuit for the time evolution step is shown in Fig. \ref{fig:MPDO_evolve} where $\Tilde{\mathcal{U}}$ are the two-qubit super-operators defined in the reweighted Pauli basis (Eq. \eqref{eq:boson_rewt}).

With $\gamma=1$, rTEBD becomes the same as TEBD time evolution of a MPDO.
The idea behind rTEBD is that when $\gamma>1$, the $A$ matrices will have to compensate for $\gamma$ by producing smaller coefficients for $A_1^{\mu_1}A_2^{\mu_2}\cdots A_L^{\mu_L}$, by a factor of $\gamma^{-n}$, when there are $n$ many non-identity Pauli matrices in $\tilde{\sigma}^{\mu_1 \cdots \mu_L}$.
The SVD truncation will then more aggressively approximate terms with more non-identity Pauli matrices, while more cautiously maintaining the accuracy of terms with less non-identity Pauli matrices. Hence, expectation values involving a small number of Pauli operators will be conserved to a higher accuracy.

The unitary super-operators also need to be written in the reweighted Pauli basis, after which the super-operators are no longer unitary. 
To do this, we first define a dual basis of Pauli operators:
\begin{equation}
    \Bar{\sigma}^{\mu} = \begin{cases}
        \sigma^0 & \text{if } \mu=0 \\
        \frac{1}{\gamma} \sigma^{\mu} & \text{if } \mu \neq 0
    \end{cases}
    \label{eq:boson_undo}
\end{equation}
which satisfy $\tr \tilde{\sigma}^\mu \Bar{\sigma}^\nu = 2 \delta^{\mu\nu}$. Throughout the text, we use the notation $\tilde{\sigma}^{\mu}$ for reweighted Pauli operators and $\bar{\sigma}^{\mu}$ for the Pauli operators in the dual basis.

In the reweighted Pauli basis, the unitary super-operators are defined as
\begin{equation}
    \mathcal{\tilde{U}}^{\nu_1\nu_2\mu_1\mu_2} = \frac{1}{4}\Tr[ \Bar{\sigma}^{\nu_1}\cdot\Bar{\sigma}^{\nu_2}\cdot U\cdot\Tilde{\sigma}^{\mu_1}\cdot\Tilde{\sigma}^{\mu_2}\cdot U^{\dagger}]
    \label{eq:U_rewt}
\end{equation}
where $U = e^{-i\Tilde{H}\delta t}$, $\delta t$ is the Trotter step, and $\Tilde{H}$ is a local $2$-qubit Hamiltonian.

We stress that the non-unitarity of the super-operator is just a mathematical formulation that does not affect the unitary behavior of the time evolution. The undoing of the reweighting preserves the unitarity of the time evolution.
Only the SVD truncation breaks unitarity.

For a product state $\rho = \otimes_i \rho_i$, where each $\rho_i$ is a single-qubit density matrix,
  the $\chi=1$ MPDO matrices in the reweighted Pauli basis are: 
\begin{equation}
    A_i^{\mu} = \Tr [\Bar{\sigma}^{\mu}\cdot\rho_i] \; \mathbb{1}_1
    \label{eq:A_rewt}
\end{equation}
where $\mathbb{1}_1$ is a $1 \times 1$ identity matrix.
We show the derivation of $A_i^{\mu}$ and $\mathcal{\tilde{U}}^{\nu_1\nu_2\mu_1\mu_2}$ in the reweighted Pauli basis in Appendix \ref{ap:rewt}.

\subsection{Fermionic systems}
To simulate fermionic systems, we make use of the Jordan-Wigner transformation to map a fermion chain to qubits \cite{Jordan1928}.
To define the Jordan-Wigner transformation, it's convenient to first map the Pauli operators to hard-core\footnote{%
  ``Hard-core'' means that a $b_j^\dagger b_j \leq 1$ constraint is applied to the boson Hilbert space.}
  boson $b_j$ operators:
\begin{align}
     \sigma^+_j &= \frac{1}{2}(\sigma^x_j + i \sigma^y_j) = b^{\dagger}_j \nonumber\\
     \sigma^-_j &= \frac{1}{2}(\sigma^x_j - i \sigma^y_j) = b_j \nonumber \\
     \sigma^z_j &= 2b^{\dagger}_j b_j - \mathbb{1}
\end{align}
The Jordan-Wigner transformation then transforms the hard-core boson operators into fermion operators $c_j$:
\begin{equation}
     c_j =  b_j \prod_{k=1}^{j-1} \sigma^z_k 
\end{equation}
which satisfy the usual $\{c_i, c_j^\dagger\} = \delta_{i,j}$ anti-commutation relations.

For simulating fermions, we choose to use a slightly different reweighted basis:
\begin{equation}
    \Tilde{\sigma}^{\mu}_{\text{F}} = \begin{cases}
        \sigma^0 & \text{if } \mu=0 \\
        \gamma \sigma^{\mu} & \text{if } \mu=x,y \\
        \gamma^2 \sigma^z & \text{if } \mu=z
    \end{cases}
    \label{eq:fermion_rewt}
\end{equation}
This basis reweights each factor of $c_j$ by $\gamma$.
Thus, $\sigma^z$ is reweighted by $\gamma^2$ because $\sigma^z$ is a product of two fermion operators.
In Appendix \ref{ap:f_rewt}, we show that this scheme of reweighted fermion operators is slightly better than the bosonic scheme in \eqref{eq:boson_rewt} for non-interacting fermion chains.
The dual Pauli operators in this basis are:
\begin{equation}
    \Bar{\sigma}^{\mu}_{\text{F}} = \begin{cases}
        \sigma^0 & \text{if } \mu=0 \\
        \frac{1}{\gamma} \sigma^{\mu} & \text{if } \mu=x,y \\
        \frac{1}{\gamma^2} \sigma^z & \text{if } \mu=z
    \end{cases}
\end{equation}
The time evolution scheme for the fermionic case remains the same as that of the bosonic case.

rTEBD has the same $\order{L\chi^3d^6}$ time complexity as MPDO-TEBD, compared to $\order{L\chi^3d^3}$ for MPS-TEBD.  The rTEBD code for different schemes as mentioned below can be found in \cite{sayak_guha_roy_2025_17479681}.

\section{Benchmarking}
\label{sec:benchmark}

We benchmark rTEBD on a free fermion system against
  MPDO-TEBD (Sec. \ref{sec:MPDO-TEBD}), and MPS-TEBD (Sec. \ref{sec:MPS-TEBD}),
  i.e. TEBD where we either time-evolve an MPDO or an MPS. Since MPDO-TEBD is rTEBD at $\gamma=1$, it provides the natural ablation for isolating the reweighting effect; MPS-TEBD provides a standard reference point. A comparison against MPO-based steppers like $W^{II}$, where reweighting could in principle be incorporated, is left for future work.
We compare our results with the exact solution, which is easily obtained for a free fermionic system by mapping the many body problem to a single particle problem.

For MPS-TEBD, the MPS is trivially normalized $\langle \psi|\psi\rangle=1$ throughout the entire simulation.
For MPDO-TEBD, $\tr \rho$ rapidly decays to zero,
  and $\tr \rho$ is approximately preserved by rTEBD, but to high accuracy with large $\chi$.
Therefore, for rTEBD and MPDO-TEBD, we plot normalized expectation values
\begin{equation}
  \langle B \rangle \stackrel{\text{(normalized)}}{=} \frac{\tr B \rho}{\tr \rho} \label{eq:normalize}
\end{equation}
for an arbitrary operator $B$.
For MPDO-TEBD, the normalized expectation values perform slightly better than unnormalized expectation values at short times, but diverge horribly at later times.
For completeness, we also compare against MPDO-TEBD expectation values that are not normalized:
\begin{equation}
  \langle B \rangle \stackrel{\text{(unnormalized)}}{=} \tr B \rho
\end{equation}

\subsection{Free fermions}
\begin{figure*}
    \centering
    \includegraphics[width=15cm]{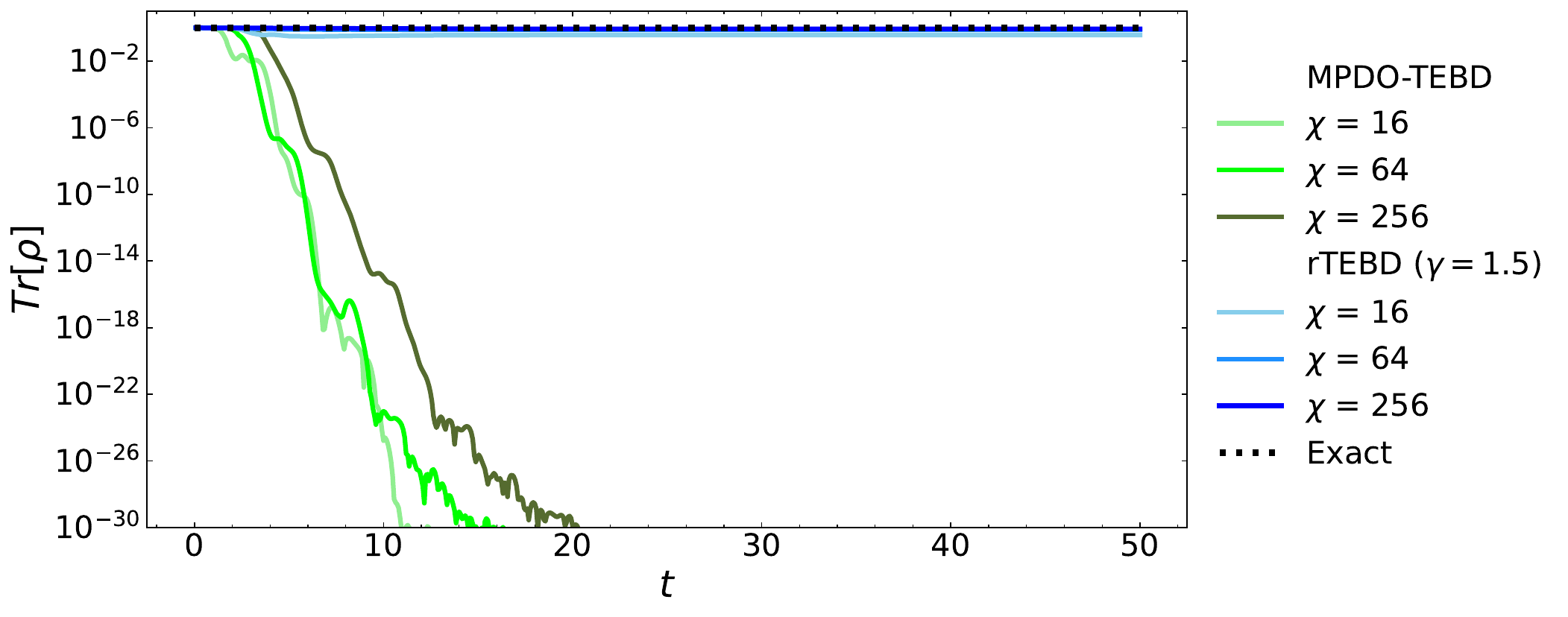}
    \caption{Plot of $\Tr[\rho]$ as a function of time on a semi-log scale for a free fermionic chain of length $L=128$ with open boundary conditions evolving according to the Hamiltonian defined in Eq. \eqref{eq:H} and starting from the initial state defined in Eq. \eqref{eq:psi_init}. We compare MPDO-TEBD with rTEBD and show that rTEBD preserves $\Tr[\rho] = 1$ approximately and gets better with increasing $\chi$. However, MPDO-TEBD is unable to preserve $\Tr[\rho]$, which decays approximately exponentially.}
    \label{fig:fermion_trace}
\end{figure*}

\begin{figure}
    \centering
    \includegraphics[width=9cm]{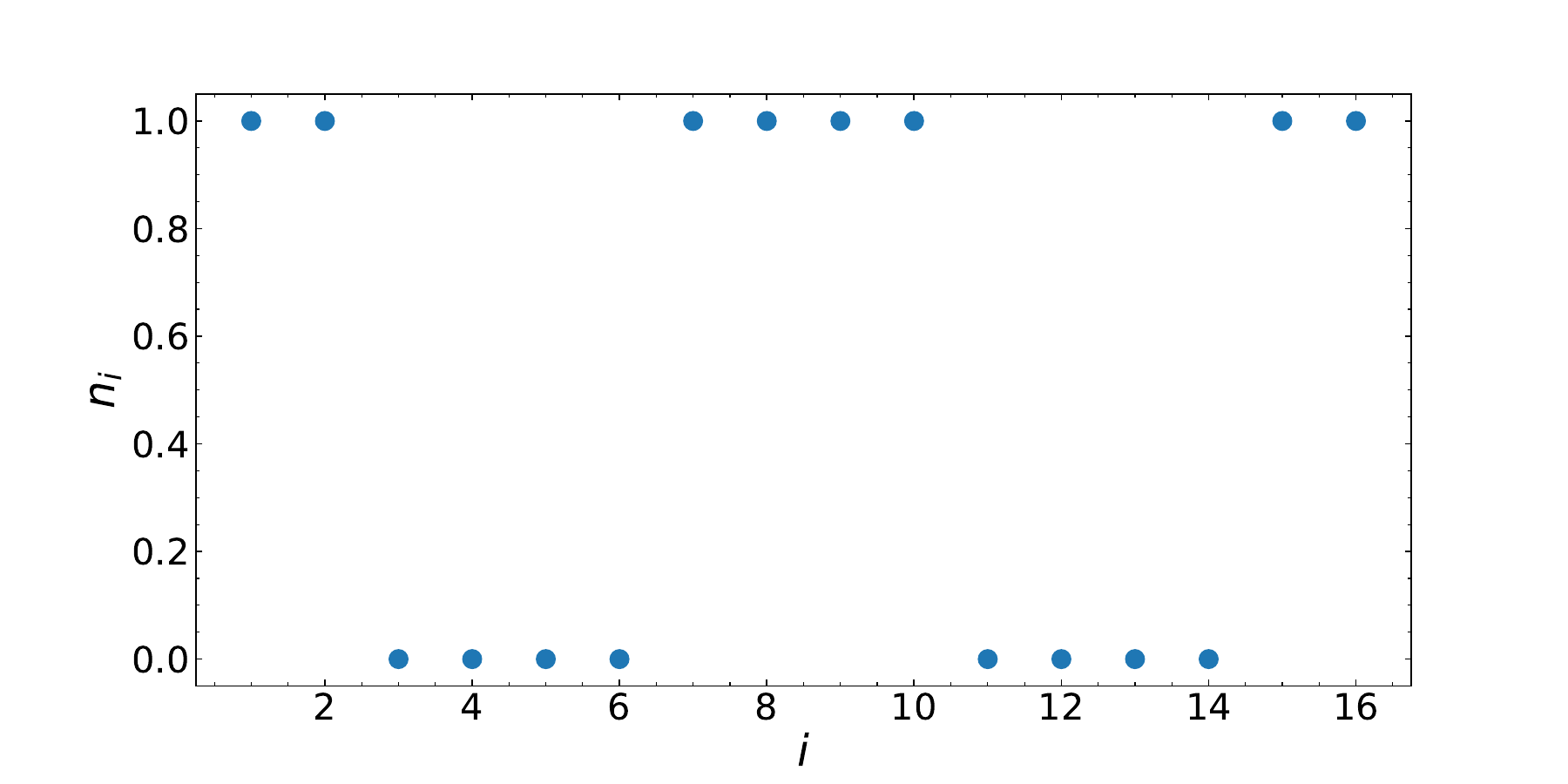}
    \caption{The initial state $\ket{\psi_0}$ [Eq. \eqref{eq:psi_init}] that we time evolve by the free-fermion chain Hamiltonian [Eq. \eqref{eq:H}].}
    \label{fig:init_state}
\end{figure}

We consider a free fermionic system defined by the Hamiltonian
\begin{align}
    H = J \sum_{\langle i,j \rangle} c_i^{\dagger}c_j + \text{h.c.}
    \label{eq:H}
\end{align}
Throughout, we use units such that $J=1$.
Following Ref \cite{White_2018}, we time evolve the following initial state:
\begin{equation}
    \ket{\psi_0} = \bigotimes_{j=1}^{L} \ket{g_j}
    \label{eq:psi_init}
\end{equation}
where
\begin{equation}
    \ket{g_j} = \begin{cases}
        \ket{1} & \text{if } j \text{ mod } 8 = 1,2,7, \text{or } 0 \\
        \ket{0} & \text{if } j \text{ mod } 8 = 3,4,5, \text{or }6
    \end{cases}
\end{equation}
which we plot in Fig. \ref{fig:init_state}. 
This state can be represented using reweighted Pauli operators using Eq. \eqref{eq:A_rewt}.

We consider a fermionic chain of length $L=128$ with open boundary conditions.
We use a Trotter step of $\delta t = 0.08$ throughout this work.
We use the same $\delta t = 0.08$ for the exact solution.
For the rTEBD technique, we use $\gamma=1.5$ as the reweighting parameter (which we found to work better than $\gamma=2$ or higher by comparing rTEBD dynamics with exact dynamics across $\gamma$ in the range [1,2]).

Fig.~\ref{fig:fermion_trace} compares the trace preservation of the two methods. After the SVD truncation of the MPDOs defined in the regular Pauli basis, $\Tr[\rho]
$ decays approximately exponentially for MPDO-TEBD, with a decay rate that decreases as $\chi$ increases. rTEBD, in contrast, approximately preserves $\Tr[\rho] = 1$ throughout the simulation, with the small deviation visible at $\chi = 16$ improving systematically at larger $\chi$.

\begin{figure*}
    \centering
    \includegraphics[width=15cm,height=17cm]{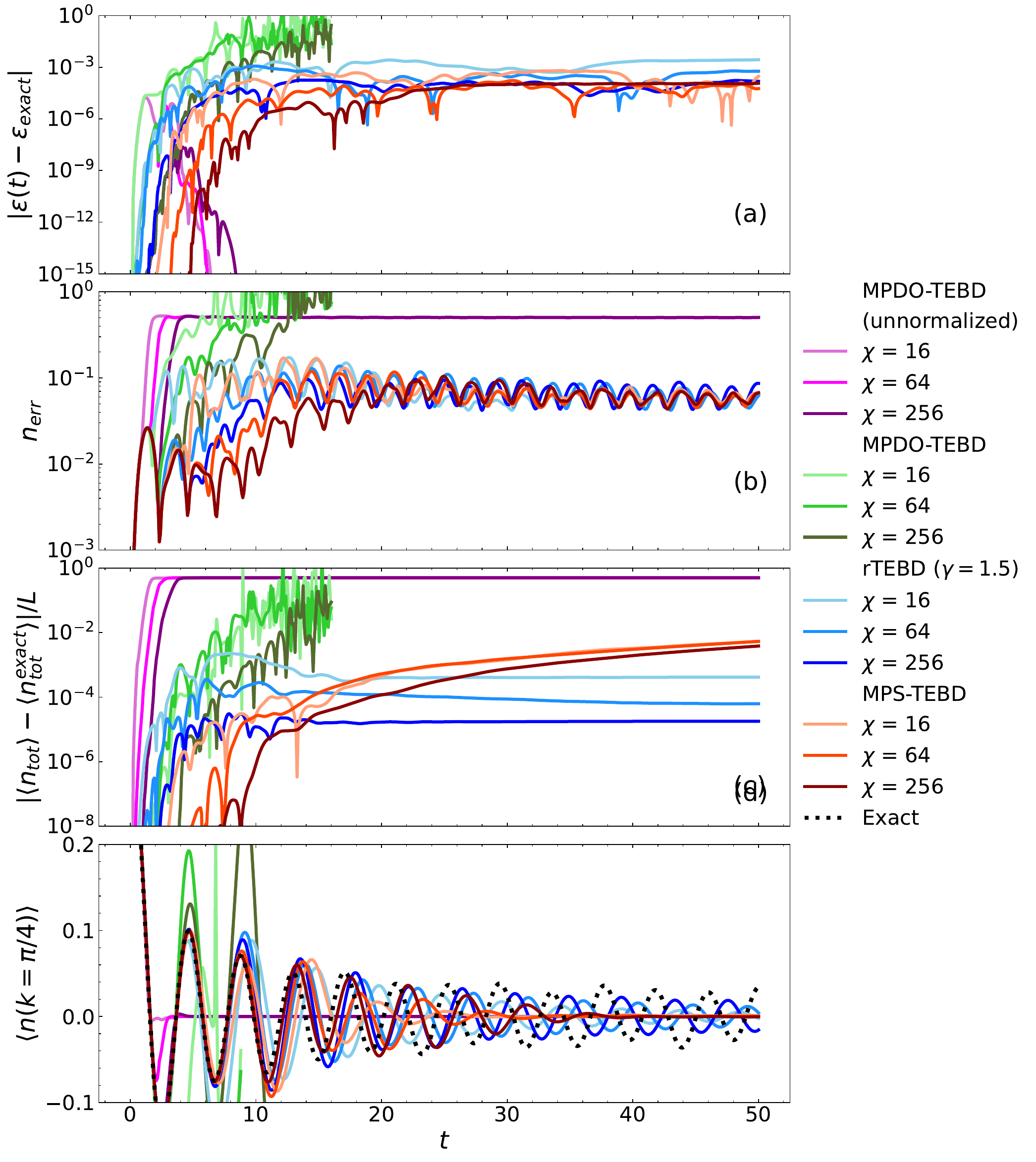}
    \caption{Plots of (a) distance to exact dynamics of the energy density ($\varepsilon$), (b) average fermion number error ($n^i_{\text{err}}$), (c) distance to exact dynamics of total fermion number ($\langle n_{\text{tot}}\rangle /L$) and (d) Fourier transform of fermion number ($\langle n(k=\pi/4) \rangle$) as a function of time, defined in Eqs. (\ref{eq:energy density}--\ref{eq:fourier}),
    for a free fermionic chain of length $L=128$ with open boundary conditions evolving according to the Hamiltonian defined in Eq. \eqref{eq:H} and starting from the initial state defined in Eq. \eqref{eq:psi_init}.
    The time evolution is performed using MPDO-TEBD (unnormalized); MPDO-TEBD, for which expectation values are normalized by the trace as in \eqref{eq:normalize}; rTEBD, our modification of MPDO-TEBD to use a reweighted Pauli basis; and MPS-TEBD, a widely used reference method. 
    Each time step is taken to be $\delta t=0.08$.
    The dotted black line shows the exact expressions for comparison.
    We show the error plots in (a), (b) and (c) in a semi-log scale while keeping a linear scale in (d). We truncate MPDO-TEBD lines as they start to diverge due to division by a vanishing $\Tr[\rho]$.}
    \label{fig:fermion_figs}
\end{figure*}

\begin{figure*}
    \includegraphics[width=12cm]{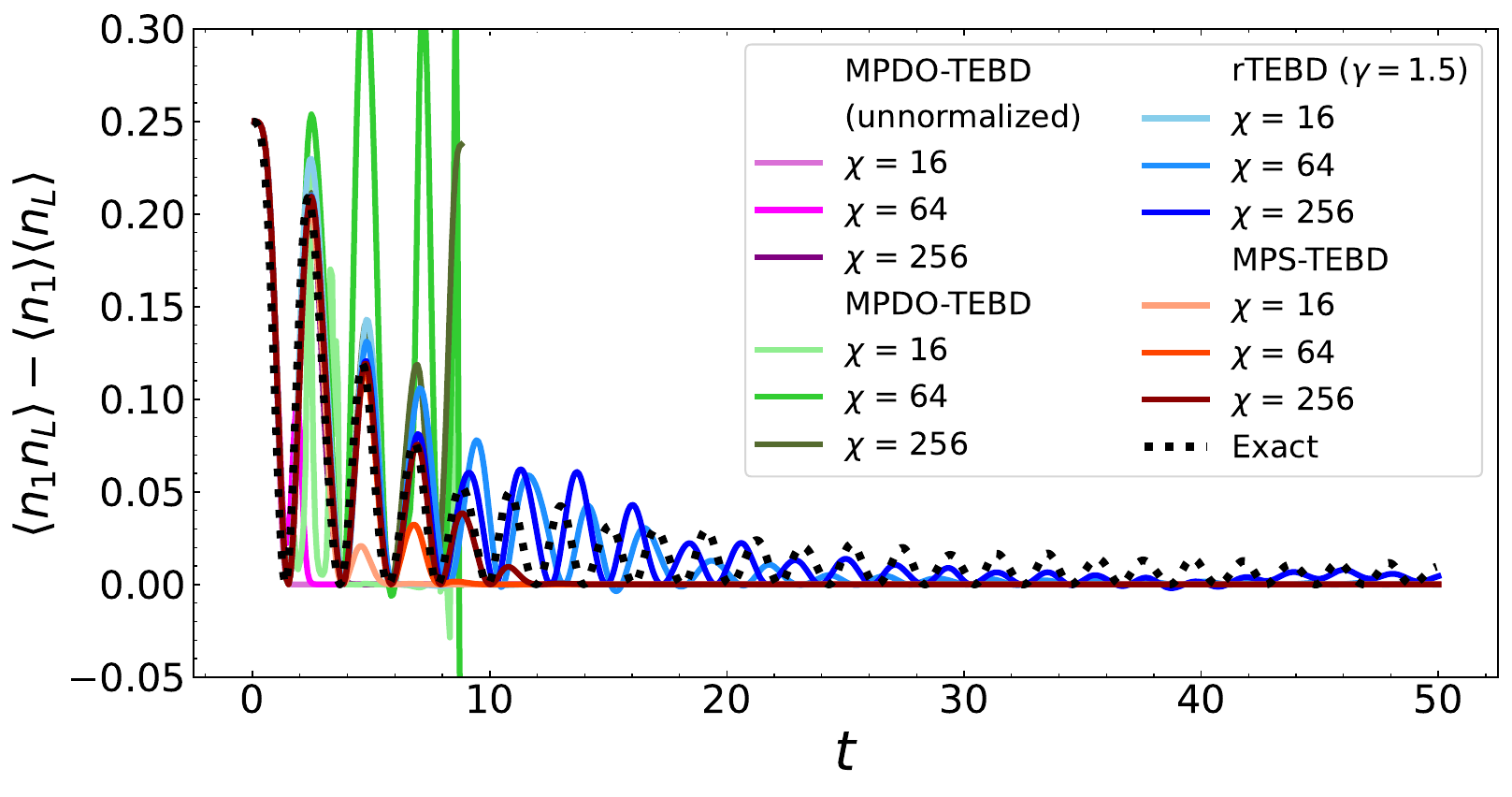}
    \caption{Simulated time-evolution of the connected density-density correlation function $\langle n_1 n_L \rangle_c$ (Eq. \eqref{eq:connected}) 
    starting from the initial GHZ state defined in Eq. \ref{eq:psi_cat}.
    Similar to Fig. \ref{fig:fermion_figs}, we simulate a free fermion chain of length $L=128$ with open boundary conditions evolving under the Hamiltonian defined in Eq. \ref{eq:H} 
    with a Trotter step $\delta t = 0.08$. 
    Similar to Fig. \ref{fig:fermion_figs}(d), we find that rTEBD most accurately preserves the decaying oscillations. We truncate MPDO-TEBD lines due to divergence.}
    \label{fig:cat_n1nL}
\end{figure*}

We then compare the algorithms using several important observables:
\begin{align}
    \epsilon &= \frac{\langle H \rangle}{L} \label{eq:energy density} \\
    n_{\text{err}} &= \sqrt{\frac{\sum_i( \langle n_i \rangle - \langle n_i \rangle_\text{exact})^2}{\sum_i( \langle n_i \rangle_\text{exact})^2}} \label{eq:nberr} \\
    \langle n_{\text{tot}} \rangle / L &= \frac{1}{L} \sum_i \langle n_i \rangle \label{eq:ntot}\\
    \langle n(k=\pi/4) \rangle &= \frac{1}{L} \sum_{j=1}^L e^{-ik(j-\frac{1}{2})} \langle n_j \rangle \label{eq:fourier}
\end{align}
where $\langle n_i \rangle$ is the fermion number approximation by the tensor network.
$\epsilon$ is the energy density
$n_{\text{err}}$ is the average error in the fermion number expectation value,
  where $\langle n_i \rangle_\text{exact}$ is computed exactly.
$\langle n_{\text{tot}} \rangle / L$ is the fermion number density.
$\langle n(k=\pi/4) \rangle$
is the Fourier transform of the fermion number at $k=\pi/4$, which matches the periodicity of the initial state.
Due to reflection symmetry, $\langle n(k=\pi/4) \rangle$ is real-valued at all times for the exact time evolution;
  for inexact-evolution, we plot the real value.

Fig. \ref{fig:fermion_figs} compares the different methods with rTEBD and the exact expression. In Fig.~\ref{fig:fermion_figs} (a) and (c), we consider the absolute error or distance to exact dynamics for the energy density and the total fermion number plotted on a semi-log scale.
For the energy density, Fig.~\ref{fig:fermion_figs} (a) shows rTEBD performs significantly better than MPDO-TEBD and roughly as well as MPS-TEBD.
The energy density calculated by MPDO-TEBD (unnormalized) is trivially very close to zero, the exact energy density, only because the density matrix decays to zero for this method. This makes MPDO-TEBD accidentally look like the best method.
When we look at the average fermion number error ($n_{\text{err}}$) plot in Fig.~\ref{fig:fermion_figs} (b) on a semi-log scale, we see a similar trend where rTEBD performs much better than MPDO-TEBD and MPDO-TEBD (unnormalized) while roughly matching the accuracy of MPS-TEBD.
When we look at the total fermion number error plot in Fig.~\ref{fig:fermion_figs} (c), we see that rTEBD conserves the total fermion number better than all the other methods including MPS-TEBD. In fact, at long times, the error for $\chi=256$ rTEBD is about two orders of magnitude better than $\chi=256$ MPS-TEBD.
In Fig.~\ref{fig:fermion_figs} (d), we plot $\langle n(k=\pi/4)\rangle$; we see that rTEBD preserves the oscillation amplitude better than MPS-TEBD throughout the simulation. Both methods accumulate a small phase drift relative to the exact result at long times, reflecting Trotter and truncation errors common to TEBD-type schemes. We retain Fig.~\ref{fig:fermion_figs} (d) on a linear scale since the quantity is an oscillating quantity. A logarithmic representation of either the value or the absolute error obscures this amplitude information through frequent zero-crossings.

In summary, rTEBD is significantly more accurate than MPDO-TEBD (with or without normalizing by the trace) and slightly more accurate than MPS-TEBD.
Since the density matrix decays to zero with MPDO-TEBD, all MPDO-TEBD expectation values decay to zero when not normalized by the trace, while all MPDO-TEBD expectation values quickly diverge when normalizing by the trace.
rTEBD, on the other hand, does a much better job at preventing the decay of the density matrix, and thus avoids both of these issues.

As mentioned in the introduction, we expect rTEBD to preserve long-range two-body correlators better than other methods. To show this, we consider the time evolution of the connected density-density correlation function $\langle n_1 n_L \rangle_c$ (again for a chain of length $L=128$):
\begin{equation}
    \langle n_1 n_L \rangle_c = \langle n_1 n_L \rangle - \langle n_1 \rangle\langle n_L \rangle \label{eq:connected}
\end{equation}

To give this correlation a large amplitude, we choose a GHZ state as the initial state (rather than the previous uncorrelated initial state):
\begin{equation}
    \ket{\psi_\text{GHZ}} = \frac{1}{\sqrt{2}}(\ket{\psi_0}+\prod_j \sigma_j^1 \ket{\psi_0})
    \label{eq:psi_cat}
\end{equation}
where $\ket{\psi_0}$ (Eq. \eqref{eq:psi_init}) was our previous uncorrelated initial state.
Fig. \ref{fig:cat_n1nL} compares the different algorithms for this evolution. It demonstrates rTEBD's ability to preserve long-range correlations. The connected correlator $\langle n_1 n_L \rangle_c$ probes correlations between the end points of the chain, precisely a kind of low-weight long-range observable that rTEBD is designed to preserve. We find that rTEBD tracks the exact result closely throughout the simulation, with systematic convergence toward the exact dynamics as $\chi$ increases. In contrast, MPS-TEBD damps the oscillation amplitude significantly faster than the exact result, with even $\chi=256$ failing to capture the long-time behavior. MPDO-TEBD fails entirely on this observable: the unnormalized version decays to zero with the trace, while the normalized version diverges as it amplifies numerical noise through division by a vanishing $\Tr[\rho]$.

\subsection{Interacting spin model}
\label{sec:spin model}

\begin{figure}
    \centering
    \includegraphics[width=8cm]{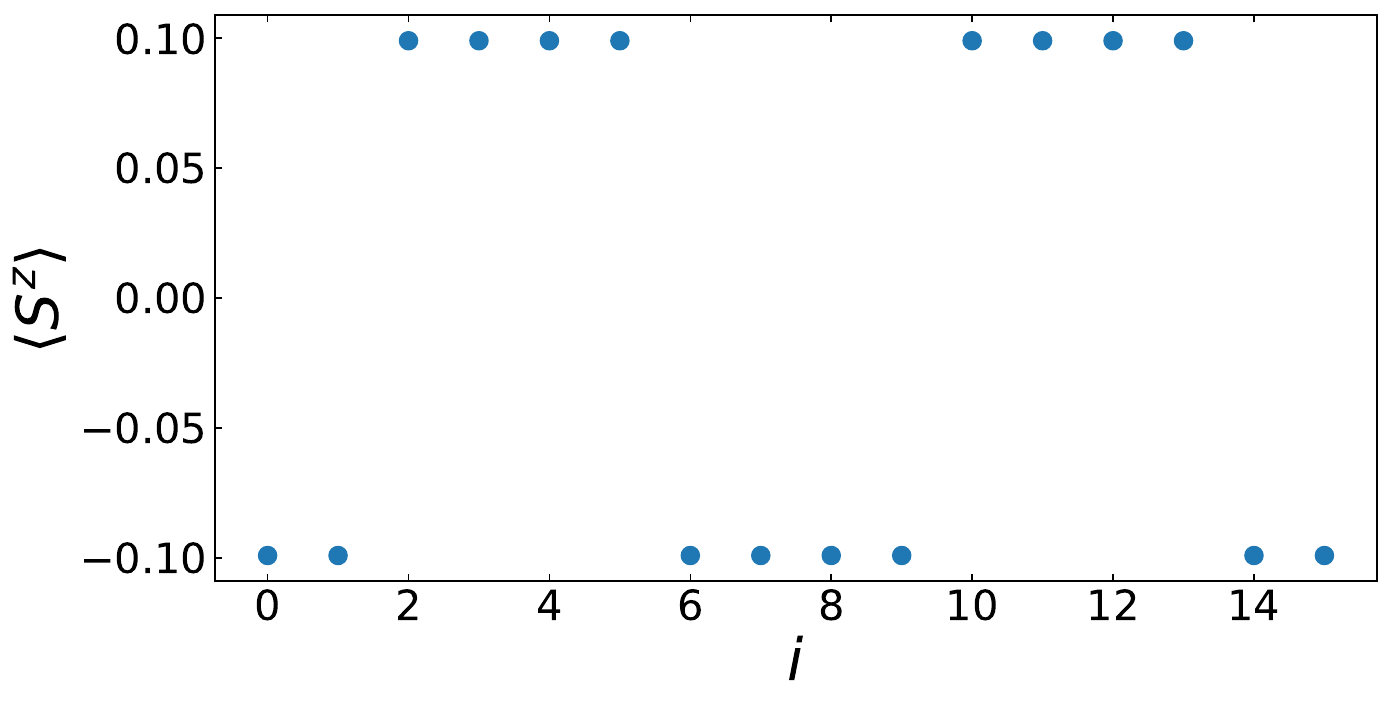}
    \caption{The initial state $\ket{\psi_0}$ [Eq. \eqref{eq:spin_init}] that we time evolve by the spin Hamiltonian [Eq. \eqref{eq:spin_H}].}
    \label{fig:spin_init}
\end{figure}

We also consider an interacting and non-integrable spin-$\frac{1}{2}$ system defined by the Hamiltonian
\begin{equation}
    H = J\sum_{i=1}^{L-1}S_i^zS_{i+1}^z + \frac{h^x}{2}\sum_i^L S^x_i + \frac{h^z}{2} \sum_i^L S_i^z
    \label{eq:spin_H}
\end{equation}
We consider the parameters
\begin{equation}
    J=1, \hspace{1mm} h^x = 0.9045, \hspace{1mm}h^z = 0.8090
\end{equation}
(same as \cite{White_2018}).
Throughout, we use units such that $J=1$.
The presence of the non-zero $h^z$ term makes this model non-integrable. The initial state that we time evolve is given by
\begin{equation}
    \ket{\psi_0} = \bigotimes_{j=1}^L [(1-g_j)\ket{\downarrow} + (1+g_j)\ket{\uparrow}]
    \label{eq:spin_init}
\end{equation}
where
\begin{equation}
    g_j = 0.1 \times \begin{cases}
        -1 & \text{if } j \text{ mod } 8 = 1,2,7, \text{or } 0 \\
        1 & \text{if } j \text{ mod } 8 = 3,4,5, \text{or }6
    \end{cases}
\end{equation}
We plot the $\langle S^z\rangle$ spin expectation value of this initial state in Fig. \ref{fig:spin_init}.

\begin{figure*}
    \centering
    \includegraphics[width=13cm]{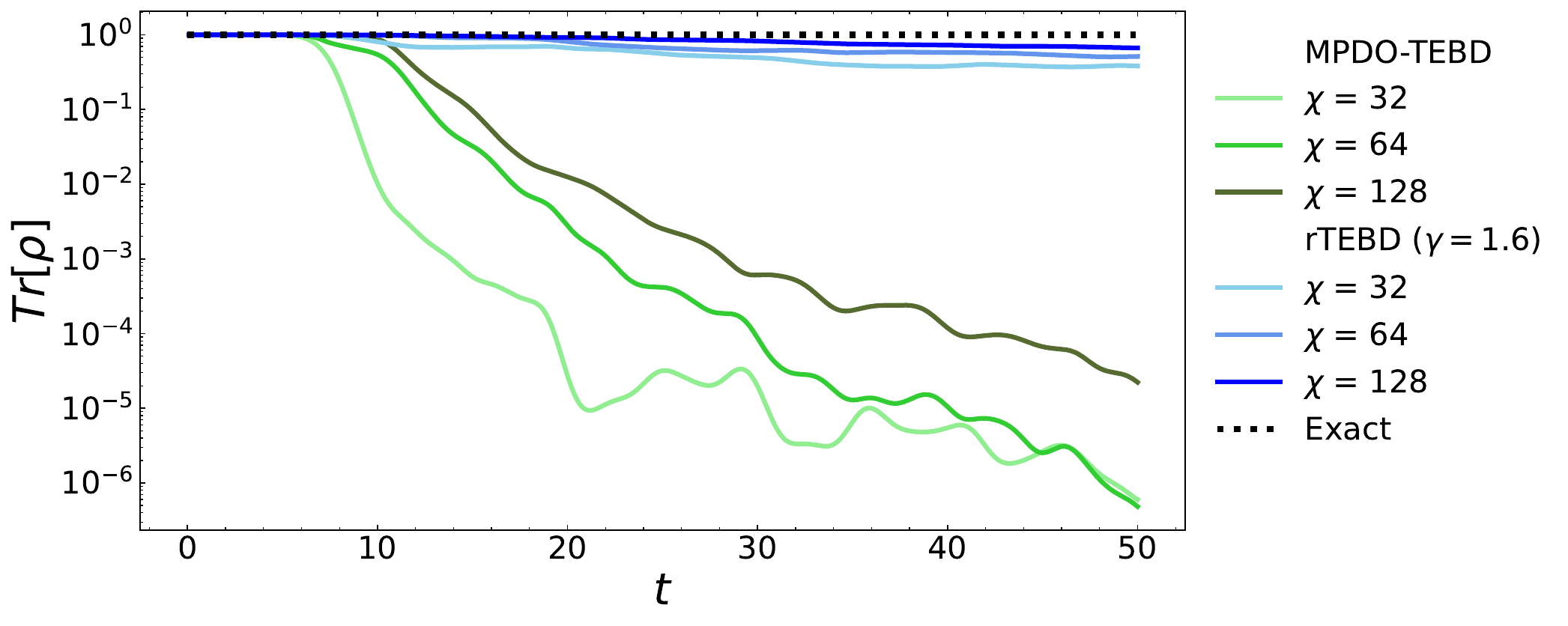}
    \caption{
    Same as Fig. \ref{fig:fermion_trace}, except for the interacting spin system defined in Sec. \ref{sec:spin model}.
    We again find that rTEBD preserves $\Tr[\rho]$ significantly better than MPDO-TEBD.
    }
    \label{fig:spin_trace}
\end{figure*}

\begin{figure*}
    \centering
    \includegraphics[width=15cm]{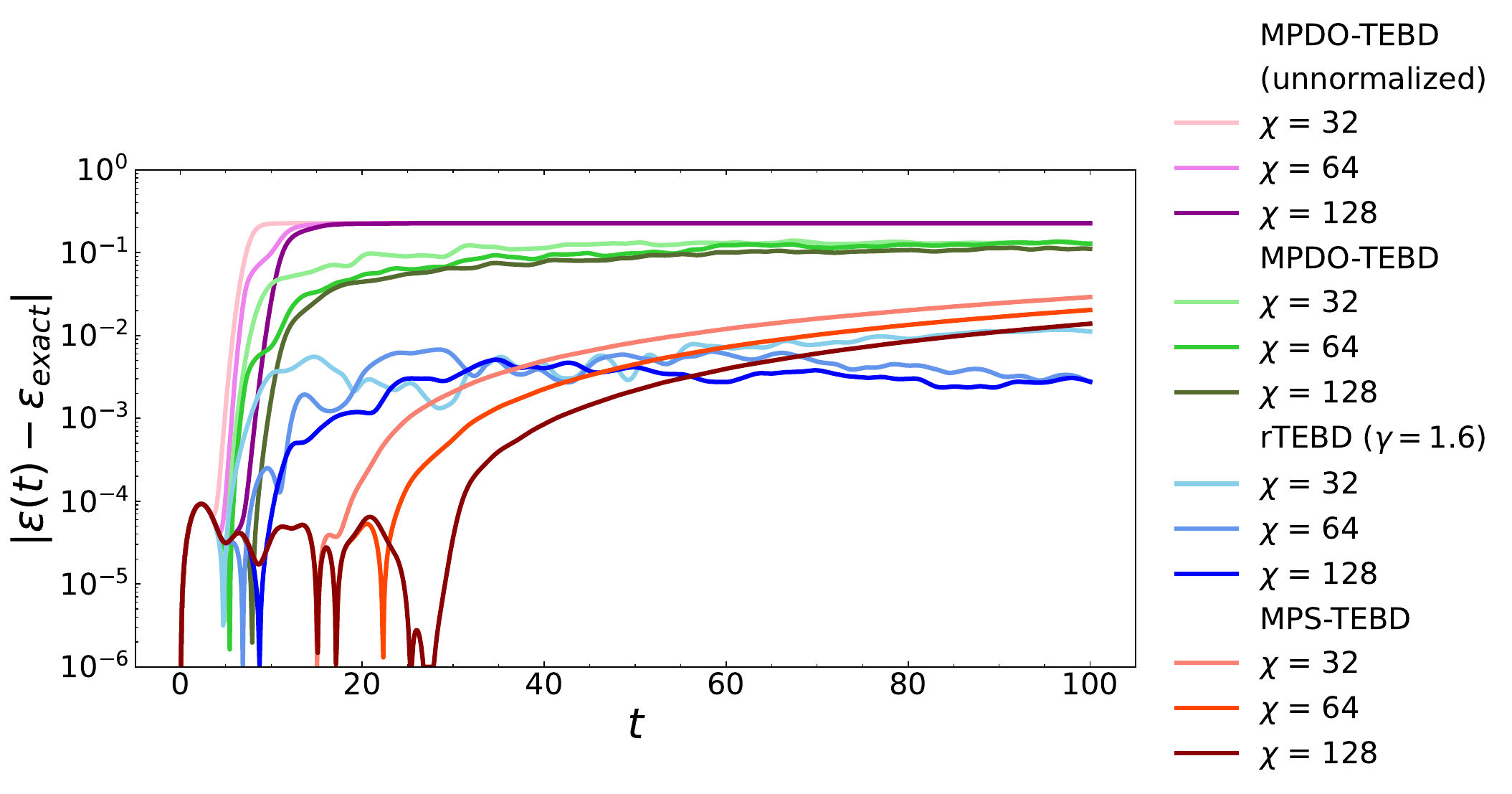}
    \caption{
    Same as Fig. \ref{fig:fermion_figs}(a), except for the interacting spin system defined in Sec. \ref{sec:spin model}.
    We find that rTEBD preserves the conserved total energy density ($\varepsilon$) significantly better than MPDO-TEBD and slightly better than MPS-TEBD.
    }
    \label{fig:spin_energy}
\end{figure*}
 Since we no longer have an exact solution, we simply study the conserved energy density for this model. In Fig.~\ref{fig:spin_energy}, we plot the absolute error of the total energy density as a function of time computed using rTEBD, MPDO-TEBD and MPS-TEBD with respect to the initial energy density ($|\varepsilon(t)-\varepsilon_{\text{exact}}|$) on a logarithmic scale. The hierarchy of methods is clearly resolved at long times: rTEBD (blue) achieves the smallest energy drift that gets better with increasing $\chi$, outperforming MPS-TEBD (red) by nearly an order of magnitude for $\chi = 128$. MPDO-TEBD (green) and unnormalized MPDO-TEBD (magenta) saturate at much larger error values, reflecting the rapid decay of $\Tr[\rho]$ (Fig.~\ref{fig:spin_trace}). At short times, MPS-TEBD has smaller error than rTEBD, while at long times rTEBD has smaller error. The crossover reflects a fundamental trade-off: MPS-TEBD efficiently represents pure states at small bond dimension but loses accuracy as entanglement grows, while rTEBD operates on the larger MPDO representation but maintains few-body operator accuracy through the reweighted truncation as entanglement grows. Similar to the fermionic case in Fig.~\ref{fig:fermion_trace}, Fig.~\ref{fig:spin_trace} shows that $\Tr[\rho]$ decays approximately exponentially for MPDO-TEBD while rTEBD approximately preserves $\Tr[\rho]=1$, which visibly gets better with increasing bond dimension $\chi$.

\subsubsection{Choosing $\gamma$}

Looking at errors in conserved quantities offers a method to tune $\gamma$.
Here, we consider using the root-mean-squared (over time) error in the conserved energy density:
\begin{equation}
    \varepsilon^\text{avg}_\text{err} = \sqrt{ \frac{1}{T_\text{f}} \int_0^{T_\text{f}} dt \, |\varepsilon(t) - \varepsilon(0)|^2 }
\end{equation}
We expect that smaller $\varepsilon^\text{avg}_\text{err}$ indicates a better choice of $\gamma$.

\begin{figure}
    \centering
    \includegraphics[width=9cm]{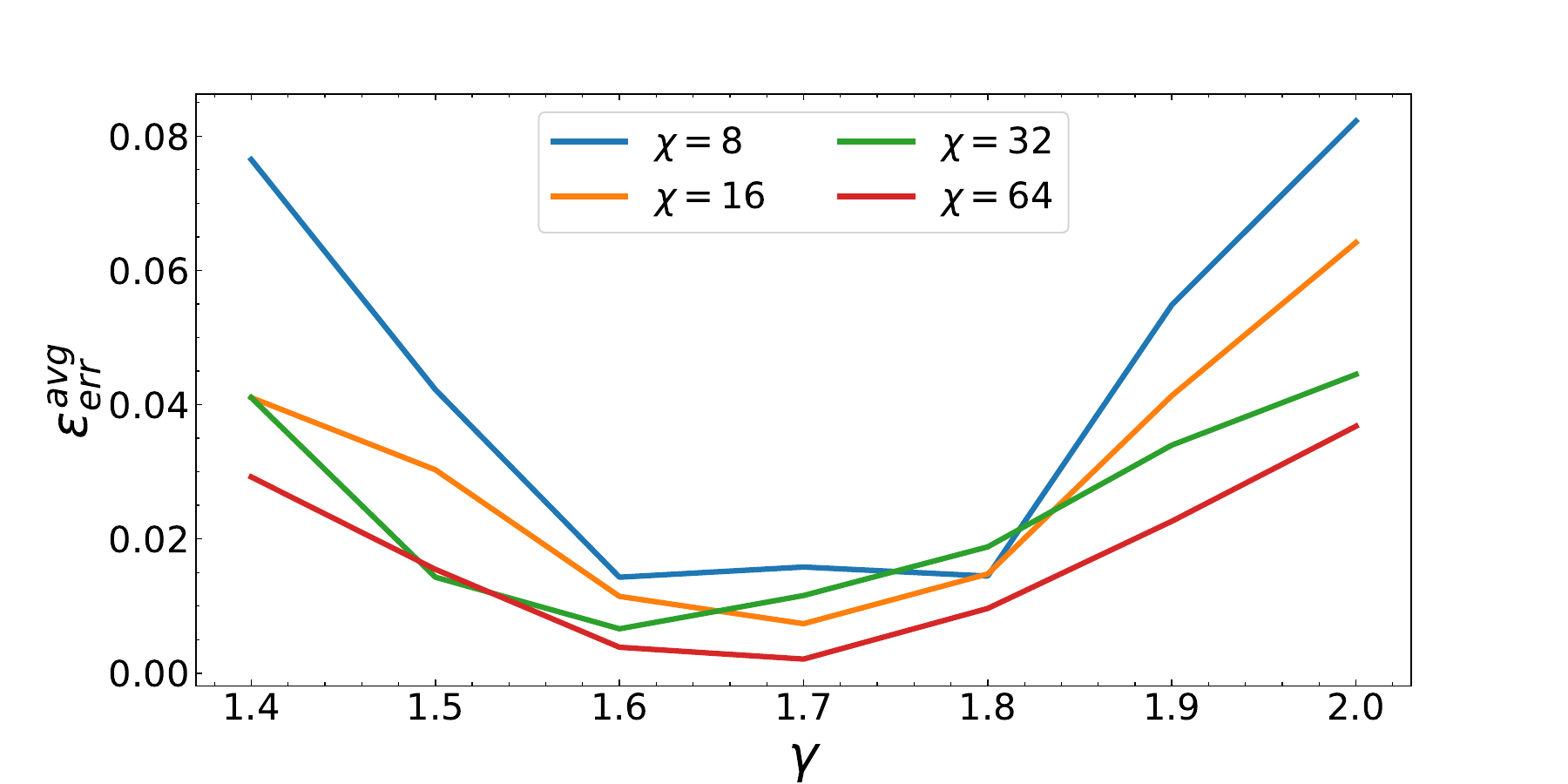}
    \caption{
    Plot of $\varepsilon^\text{avg}_\text{err}$ with $T_\text{f}=100$ as a function of $\gamma$ for different $\chi$
    and for the spin system described in Sec. \ref{sec:spin model} of length $L=64$.
    The plot suggests that values of $\gamma$ between 1.6 and 1.7 are the most accurate for this model.}    
    \label{fig:gamma_test}
\end{figure}

In Fig. \ref{fig:gamma_test} we plot $\varepsilon^\text{avg}_\text{err}$ with $T_\text{f} = 100$ in units of $J$ as a function of $\gamma$ and for a range of $\chi$ for the interacting spin model (Eq.~\ref{eq:spin_H}).
The plot suggests that values of $\gamma$ between 1.6 and 1.7 are the most accurate for this model. We choose $\gamma = 1.6$ for Fig \ref{fig:spin_energy} because $\gamma=1.6$ has lower $\varepsilon^\text{avg}_\text{err}$ going up to time $t=100$.

 The increase in the error with increasing $\gamma$ likely results from a poor balance between preserving low-weight expectation values while not completely ignoring higher-weight expectation values.
 We leave a better understanding of these effects to future work.

\section{Conclusion}
\label{sec:conc}

We introduce a new quantum many-body time evolution algorithm, Reweighted Time-Evolving Block Decimation (rTEBD).
rTEBD improves upon TEBD by using a reweighted basis of density matrices, which causes the SVD truncation step to more accurately preserve the expectation value of few-body operators by deprioritizing the accuracy of correlation functions involving the product of many operators.

We benchmark rTEBD on large, time-evolving one-dimensional free fermionic chain, where we can compare against exact results, and bosonic models, for which we can exactly calculate the error for conserved quantities.
We compare rTEBD with TEBD of matrix product density operators (MPDO) and matrix product states (MPS).
From Figs. \ref{fig:fermion_figs} and \ref{fig:cat_n1nL}, we find that rTEBD preserves conserved quantities, oscillations, and long-range correlations better than MPDO-TEBD and MPS-TEBD.
Hence, we show that a matrix reweighting technique used on matrix product density operators can improve quantum dynamics simulations.

rTEBD involves a reweighting factor $\gamma>1$, where $\gamma=1$ reduces rTEBD to the prior MPDO-TEBD method.
We found that values of $\gamma$ near $\gamma=1.5$ perform well for the fermionic case and around $\gamma=1.6$ for the spin model. 
To systematically choose a good value of $\gamma$ for systems with at least one conserved quantity,
we suggest sweeping across a range of $\gamma$ and choose the $\gamma$ with the least error for the conserved quantity. 
It would be interesting to develop a better theoretical understanding of the effects of large $\gamma$.

Fermionic single-particle correlators such as $G_{ij} = \langle c_i^{\dagger}c_j\rangle$ or the momentum number operator $n_k$ involve nonlocal Jordan-Wigner parity strings when expressed in a Pauli operator basis. As a consequence, reweighted truncation schemes that bias against high-weight operator content can suppress long-range fermionic coherence, even when local observables are well converged. This effect is illustrated in Appendix \ref{ap:f_rewt}, where different reweighting schemes show markedly different behavior for $n_k(t)$. Developing reweighting strategies that preserve fermionic coherence while retaining the advantages of rTEBD is an important direction for future work.

Additionally, we plan to compare rTEBD with existing time evolution methods, such as DMT \cite{White_2018}, DAOE \cite{DAOE}, LITE \cite{local-info1}, OST dynamics \cite{PhysRevB.110.134308}, sparse Pauli dynamics \cite{sparse} and variational methods \cite{PRXQuantum.5.020361,Bauernfeind_Aichhorn_2019,Goto_Danshita_2019,Haegeman_Cirac_Osborne_Pižorn_Verschelde_Verstraete_2011}. 
Extending the rTEBD algorithm to studying open system dynamics, imaginary time dynamics, phase transitions, and ground states is also left for future work. Integrating the reweighting idea of rTEBD into existing MPO-based methods like $W^{II}$ steppers \cite{Wtwo_stepper} is also an interesting research direction.

\section*{Acknowledgements}
We would greatly like to thank Kaden Hazzard, Vaibhav Sharma, Haotian Wei, Christopher White and Jonathan Stepp for invaluable insights. We would also like to thank Annabelle Bohrdt, Ayushi Singhania, Pranay Patil, Oliver Dudgeon, and Matej Moško for useful conversations. 

\paragraph{Funding information}
This research was supported in part by the Welch Foundation through Grant No. C-2166-20230405. For running our simulations, this work was supported in part by the Big-Data Private-Cloud Research Cyberinfrastructure MRI-award funded by NSF under grant CNS-1338099 and by Rice University's Center for Research Computing (CRC).

\appendix

\section{Derivation of the MPDOs and unitaries in the reweighted basis}
\label{ap:rewt}

In this section, we will derive Eq. \ref{eq:A_rewt} and the time-evolving two-qubit unitaries (Eq. \ref{eq:U_rewt}) in the reweighted Pauli basis. 

To see Eq. \ref{eq:A_rewt}, we can consider a 1-qubit system whose density operator is
\begin{equation}
    \rho = \frac{1}{2} \sum_{\mu} \Tilde{\sigma}^{\mu}A^{\mu}
    \label{eq:rho_1q}
\end{equation}
Here, $\Tilde{\sigma}^{\mu}$ are the reweighted Pauli matrices defined in Eq. \ref{eq:boson_rewt}. Using the definition of $\Bar{\sigma}^{\mu}$ in Eq. \ref{eq:boson_undo}, one can prove the following identity
\begin{equation}
    \frac{1}{2}\sum_{\mu} \Tr[\Bar{\sigma}^{\mu}\rho]\Tilde{\sigma}^{\mu} = \rho
    \label{eq:id_1}
\end{equation}
The above identity is basically applying a linear operator on $\rho$, which is seen to be the identity operator. Comparing Eqs. \ref{eq:rho_1q} and \ref{eq:id_1}, we see that
\begin{equation}
    A^{\mu} = \Tr[\Bar{\sigma}^{\mu}\rho]
\end{equation}
The mapping to many qubit system is straightforward by adding the relevant position indices. 

Similar math can be used to derive the unitaries in the Pauli basis. For this case, we consider a 2-qubit system with density operator in the reweighted basis
\begin{equation}
    \rho = \frac{1}{2^2}\sum_{\mu_1,\mu_2}\Tilde{\sigma}^{\mu_1} \otimes \Tilde{\sigma}^{\mu_2} A_1^{\mu_1}A_2^{\mu_2}
\end{equation}
For time evolution of $\rho$, we apply the time evolution unitary operator $U=e^{-iH \delta t}$ and we obtain
\begin{align}
    U\rho U^{\dagger} &= \frac{1}{4}\sum_{\mu_1,\mu_2}U\Tilde{\sigma}^{\mu_1} \otimes \Tilde{\sigma}^{\mu_2}U^{\dagger} A_1^{\mu_1}A_2^{\mu_2} \nonumber\\
    &= \sum_{\mu_1,\mu_2}\sum_{\nu_1,\nu_2} \Tr_{12}[\Bar{\sigma}_1^{\nu_1}\Bar{\sigma}_2^{\nu_2}U\Tilde{\sigma_1}^{\mu_1}\Tilde{\sigma_2}^{\mu_2}U^{\dagger}]\Tilde{\sigma_1}^{\nu_1}\Tilde{\sigma_2}^{\nu_2}A_1^{\mu_1}A_2^{\mu_2}
\end{align}
The second line of the above identity is again similar to applying a linear operator that acts as the identity. Hence, we can write the unitary in the reweighted Pauli basis as
\begin{equation}
    \mathcal{\tilde{U}}^{\nu_1\nu_2\mu_1\mu_2} = \frac{1}{4}\Tr[ \Bar{\sigma}^{\nu_1}\cdot\Bar{\sigma}^{\nu_2}\cdot U\cdot\Tilde{\sigma}^{\mu_1}\cdot\Tilde{\sigma}^{\mu_2}\cdot U^{\dagger}]
\end{equation}

\section{Reweighting scheme for fermions}
\label{ap:f_rewt}
For fermionic systems, we reweight the fermionic operators, mapped to spin operators, using the following scheme. For the purpose of comparisons, we name this \emph{fermionic scheme}.
\begin{equation}
    \Tilde{\tau}^{\mu} = \begin{cases}
        \tau^0 & \text{if } \mu=0 \\
        \gamma \tau^{\mu} & \text{if } \mu=x,y \\
        \gamma^2 \tau^z & \text{if } \mu=z
    \end{cases}
    \label{eq:fermion_rewt2}
\end{equation}
We show in this appendix that the scheme in \eqref{eq:fermion_rewt2} is a better scheme compared to
\begin{equation}
    \Tilde{\tau}^{\mu} = \begin{cases}
        \tau^0 & \text{if } \mu=0 \\
        \gamma \tau^{\mu} & \text{if } \mu \neq 0
    \end{cases}
    \label{eq:fermion_rewt3}
\end{equation}
We name the scheme shown in \eqref{eq:fermion_rewt3} \emph{bosonic scheme}. We plot the energy density ($\epsilon$), the total fermion number density ($\langle n_{\text{tot}}\rangle /L$) and the average fermion number ($n_{\text{err}}$) as a function of time for the two schemes.

\begin{figure*}
    \centering
    \includegraphics[width=15cm]{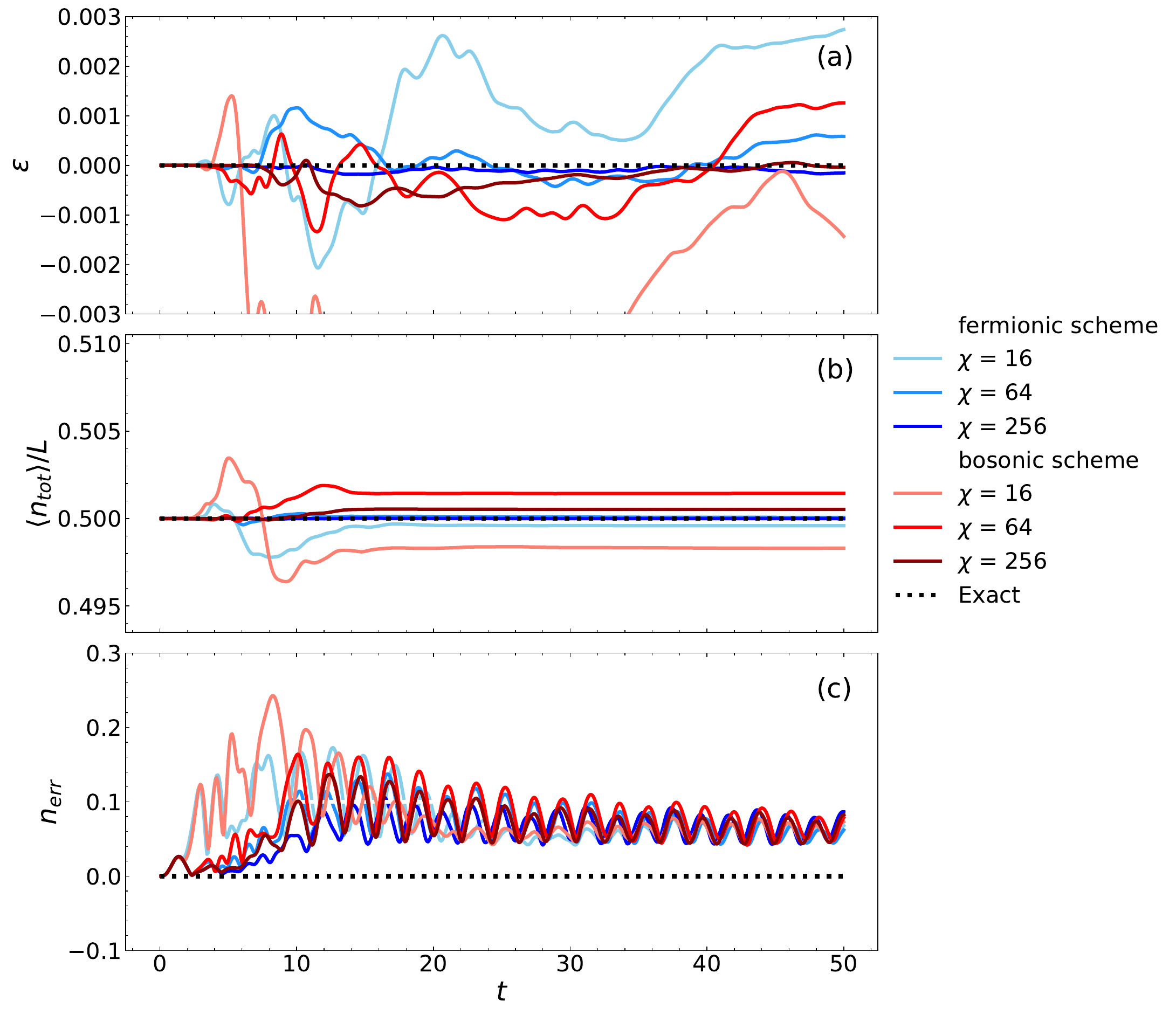}
    \caption{Plot of (a) energy density ($\epsilon$), (b) total fermion number density ($\langle n_{\text{tot}}\rangle /L$) and (c) average fermion number error ($n_{\text{err}}$) as a function of time for a chain of length $L=128$ for a free fermionic system described by the Hamiltonian in Eq. \eqref{eq:H} and time-evolving the initial state defined by Eq. \eqref{eq:psi_init}.
    The plot is made for two different reweighting schemes: \emph{fermionic} scheme (Eq. \eqref{eq:fermion_rewt2}) and \emph{bosonic} scheme (Eq. \eqref{eq:fermion_rewt3}). We see that the \emph{fermionic} scheme is more accurate than the \emph{bosonic} scheme.}
    \label{fig:check1}
\end{figure*}

Fig. \ref{fig:check1} shows that \emph{fermionic scheme} for reweighting the fermion operators is more accurate than \emph{bosonic scheme}, which is most pronounced in the first two plots.
Therefore, for our analysis and comparison with other time evolution simulation techniques like MPS-TEBD and MPDO-TEBD, we use \emph{fermionic scheme} for reweighting the Pauli operators of the fermion model.

We now plot fermionic correlations of the form $\langle c_i^{\dagger}c_j\rangle$. In one dimension, such correlators map to nonlocal Pauli strings via the Jordan–Wigner transformation, involving a string of $\tau^z$ operators between sites $i$ and $j$. As a consequence, any reweighting scheme that biases truncation based on operator weight will also act on these parity strings, making fermionic observables a particularly sensitive diagnostic of reweighted truncation.

We first introduce a different reweighting scheme, the \emph{xy} scheme, where we do not reweight the $\tau^z$ operator,
\begin{equation}
    \Tilde{\tau}^{\mu} = \begin{cases}
        \tau^0 & \text{if } \mu=0 \\
        \gamma \tau^{\mu} & \text{if } \mu=x,y \\
         \tau^z & \text{if } \mu=z
    \end{cases}
    \label{eq:fermion_xy}
\end{equation}
In this reweighting scheme, the Jordan–Wigner string of $\tau^z$ operators is no longer directly reweighted. This choice is motivated by the observation that the $\tau^z$ string encodes fermionic anticommutation rather than physical many-body correlations.

\begin{figure*}
    \centering
    \includegraphics[width=0.8\linewidth]{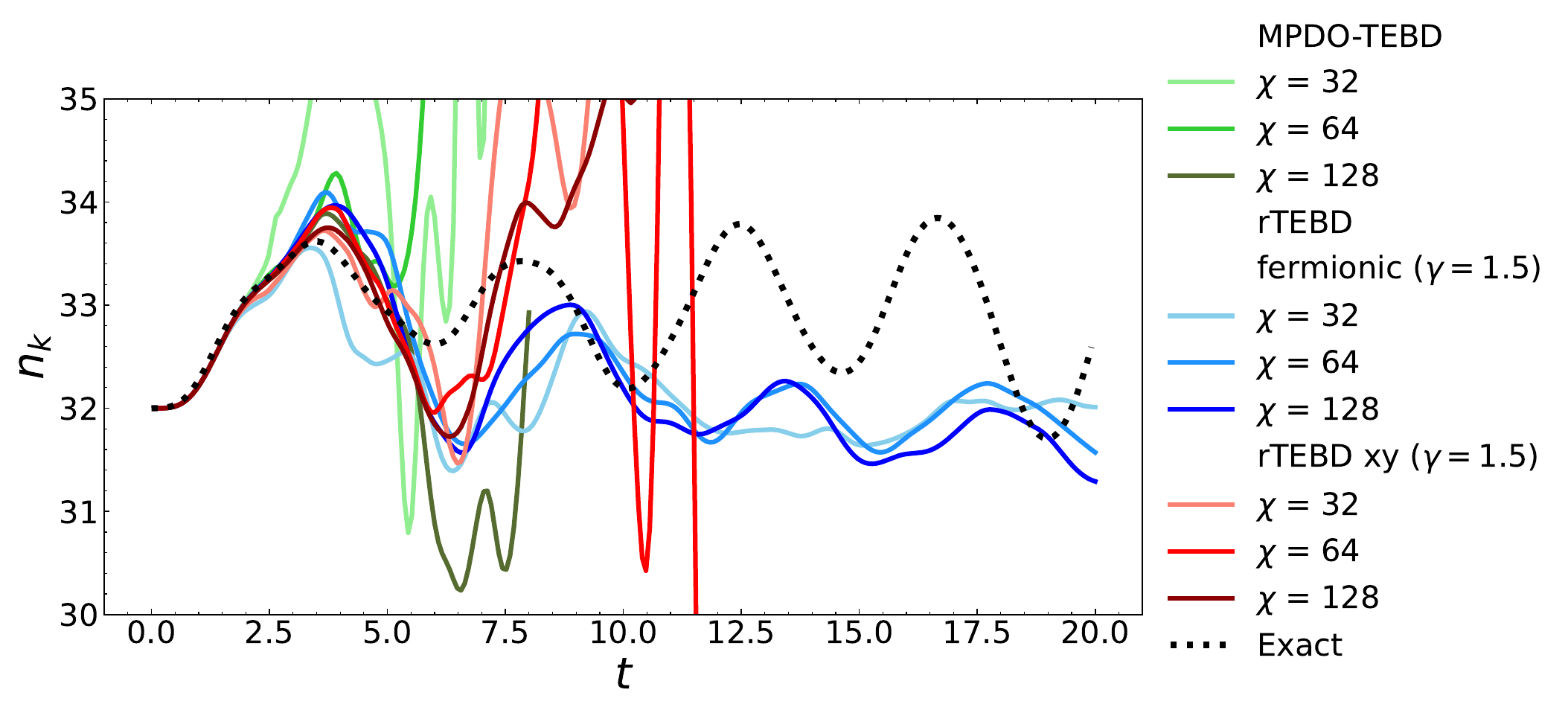}
    \caption{Plot of $n_k$ (Eq.~\ref{eq:n_k}) with $k=\pi/4$ as a function of time for a chain of length $L=64$ for a free fermionic system described by the Hamiltonian in Eq.~\eqref{eq:H}, starting from the initial state defined in Eq.~\ref{eq:psi_init}. We compare rTEBD ($\gamma=1.5$) with MPDO-TEBD and exact results. The \emph{fermionic} reweighting scheme significantly outperforms both MPDO-TEBD and the \emph{xy} scheme, while the latter shows substantial deviations already at short times.}
    \label{fig:correlations}
\end{figure*}

In Fig.~\ref{fig:correlations}, we plot the momentum distribution
\begin{equation}
    n_k(t) = \sum_{i,j} e^{-i k (i-j)}\langle c_i^{\dagger}(t)c_j(t)\rangle
    \label{eq:n_k}
\end{equation}
which provides a compact Fourier-space diagnostic of fermionic coherence and is directly sensitive to long-range contributions in $\langle c_i^{\dagger}c_j\rangle$.

We compute fermionic correlations $\langle c_i^{\dagger}(t)c_j(t) \rangle$ using rTEBD (both the \emph{fermionic} and \emph{xy} reweighting schemes), MPDO-TEBD, and also compute the `Exact' results for comparison, for a free-fermionic chain of length $L=64$. We consider the same initial state as in the main text.

In Fig.~\ref{fig:correlations}, we see that the \emph{fermionic} scheme outperforms the \emph{xy} scheme and MPDO-TEBD for small $\chi$. At the same time, the sensitivity of $n_k(t)$ to the reweighting choice highlights that fermionic coherence probes operator sectors that are more challenging for reweighted truncation schemes. Developing reweighting strategies that preserve fermionic correlations more faithfully remains an interesting direction for future work.

\bibliography{reweighted_tebd}

\begin{thebibliography}{10}
\providecommand{\url}[1]{\texttt{#1}}
\providecommand{\urlprefix}{URL }
\expandafter\ifx\csname urlstyle\endcsname\relax
  \providecommand{\doi}[1]{doi:\discretionary{}{}{}#1}\else
  \providecommand{\doi}{doi:\discretionary{}{}{}\begingroup \urlstyle{rm}\Url}\fi
\providecommand{\eprint}[2][]{\url{#2}}

\bibitem{Vidal1}
G.~Vidal,
\newblock \emph{Efficient simulation of one-dimensional quantum many-body systems},
\newblock Phys. Rev. Lett. \textbf{93}, 040502 (2004),
\newblock \doi{10.1103/PhysRevLett.93.040502}.

\bibitem{MPS_brickwork}
S.~Paeckel, T.~Köhler, A.~Swoboda, S.~R. Manmana, U.~Schollwöck and C.~Hubig,
\newblock \emph{Time-evolution methods for matrix-product states},
\newblock Annals of Physics \textbf{411}, 167998 (2019),
\newblock \doi{10.1016/j.aop.2019.167998}.

\bibitem{local-info1}
T.~K. Kvorning, L.~Herviou and J.~H. Bardarson,
\newblock \emph{{Time-evolution of local information: thermalization dynamics of local observables}},
\newblock SciPost Phys. \textbf{13}, 080 (2022),
\newblock \doi{10.21468/SciPostPhys.13.4.080}.

\bibitem{local-info2}
C.~Artiaco, C.~Fleckenstein, D.~Aceituno~Ch\'avez, T.~K. Kvorning and J.~H. Bardarson,
\newblock \emph{Efficient large-scale many-body quantum dynamics via local-information time evolution},
\newblock PRX Quantum \textbf{5}, 020352 (2024),
\newblock \doi{10.1103/PRXQuantum.5.020352}.

\bibitem{White_2018}
C.~D. White, M.~Zaletel, R.~S.~K. Mong and G.~Refael,
\newblock \emph{Quantum dynamics of thermalizing systems},
\newblock Physical Review B \textbf{97}(3) (2018),
\newblock \doi{10.1103/physrevb.97.035127}.

\bibitem{PhysRevLett.125.030601}
B.~Ye, F.~Machado, C.~D. White, R.~S.~K. Mong and N.~Y. Yao,
\newblock \emph{Emergent hydrodynamics in nonequilibrium quantum systems},
\newblock Phys. Rev. Lett. \textbf{125}, 030601 (2020),
\newblock \doi{10.1103/PhysRevLett.125.030601}.

\bibitem{DAOE}
T.~Rakovszky, C.~W. von Keyserlingk and F.~Pollmann,
\newblock \emph{Dissipation-assisted operator evolution method for capturing hydrodynamic transport} (2020), \eprint{2004.05177}.

\bibitem{srivatsa2024}
N.~S. Srivatsa, O.~Lunt, T.~Rakovszky and C.~von Keyserlingk,
\newblock \emph{Probing hydrodynamic crossovers with dissipation-assisted operator evolution} (2024), \eprint{2408.08249}.

\bibitem{lloyd2023}
J.~Lloyd, T.~Rakovszky, F.~Pollmann and C.~von Keyserlingk,
\newblock \emph{The ballistic to diffusive crossover in a weakly-interacting fermi gas} (2023), \eprint{2310.16043}.

\bibitem{kuo2023}
E.-J. Kuo, B.~Ware, P.~Lunts, M.~Hafezi and C.~D. White,
\newblock \emph{Energy diffusion in weakly interacting chains with fermionic dissipation-assisted operator evolution} (2023), \eprint{2311.17148}.

\bibitem{sparse}
T.~Begušić and G.~K.-L. Chan,
\newblock \emph{Real-time operator evolution in two and three dimensions via sparse pauli dynamics} (2024), \eprint{2409.03097}.

\bibitem{PhysRevB.110.134308}
S.~Yi-Thomas, B.~Ware, J.~D. Sau and C.~D. White,
\newblock \emph{Comparing numerical methods for hydrodynamics in a one-dimensional lattice spin model},
\newblock Phys. Rev. B \textbf{110}, 134308 (2024),
\newblock \doi{10.1103/PhysRevB.110.134308}.

\bibitem{PhysRevX.9.041017}
D.~E. Parker, X.~Cao, A.~Avdoshkin, T.~Scaffidi and E.~Altman,
\newblock \emph{A universal operator growth hypothesis},
\newblock Phys. Rev. X \textbf{9}, 041017 (2019),
\newblock \doi{10.1103/PhysRevX.9.041017}.

\bibitem{PhysRevB.105.245101}
C.~von Keyserlingk, F.~Pollmann and T.~Rakovszky,
\newblock \emph{Operator backflow and the classical simulation of quantum transport},
\newblock Phys. Rev. B \textbf{105}, 245101 (2022),
\newblock \doi{10.1103/PhysRevB.105.245101}.

\bibitem{PhysRevB.107.094311}
C.~D. White,
\newblock \emph{Effective dissipation rate in a liouvillian-graph picture of high-temperature quantum hydrodynamics},
\newblock Phys. Rev. B \textbf{107}, 094311 (2023),
\newblock \doi{10.1103/PhysRevB.107.094311}.

\bibitem{Hauschild_Leviatan_Bardarson_Altman_Zaletel_Pollmann_2018}
J.~Hauschild, E.~Leviatan, J.~H. Bardarson, E.~Altman, M.~P. Zaletel and F.~Pollmann,
\newblock \emph{Finding purifications with minimal entanglement},
\newblock Physical Review B \textbf{98}(23), 235163 (2018),
\newblock \doi{10.1103/PhysRevB.98.235163},
\newblock ArXiv: 1711.01288.

\bibitem{PRXQuantum.5.020361}
H.~Kim, M.~Fishman and D.~Sels,
\newblock \emph{Variational adiabatic transport of tensor networks},
\newblock PRX Quantum \textbf{5}, 020361 (2024),
\newblock \doi{10.1103/PRXQuantum.5.020361}.

\bibitem{Bauernfeind_Aichhorn_2019}
D.~Bauernfeind and M.~Aichhorn,
\newblock \emph{Time dependent variational principle for tree tensor networks},
\newblock arXiv:1908.03090 [cond-mat]  (2019),
\newblock ArXiv: 1908.03090.

\bibitem{Goto_Danshita_2019}
S.~Goto and I.~Danshita,
\newblock \emph{Performance of the time-dependent variational principle for matrix product states in long-time evolution of a pure state},
\newblock Physical Review B \textbf{99}(5), 054307 (2019),
\newblock \doi{10.1103/PhysRevB.99.054307},
\newblock ArXiv: 1809.01400.

\bibitem{Haegeman_Cirac_Osborne_Pižorn_Verschelde_Verstraete_2011}
J.~Haegeman, J.~I. Cirac, T.~J. Osborne, I.~Pižorn, H.~Verschelde and F.~Verstraete,
\newblock \emph{Time-dependent variational principle for quantum lattices},
\newblock Physical Review Letters \textbf{107}(7), 070601 (2011),
\newblock \doi{10.1103/PhysRevLett.107.070601}.

\bibitem{tangent-space-krylov}
O.~Kovalska, J.~von Delft and A.~Gleis,
\newblock \emph{Tangent space krylov computation of real-frequency spectral functions: Influence of density-assisted hopping on 2d mott physics} (2025), \eprint{2510.07279}.

\bibitem{DMRG-1}
S.~R. White,
\newblock \emph{Density matrix formulation for quantum renormalization groups},
\newblock Phys. Rev. Lett. \textbf{69}, 2863 (1992),
\newblock \doi{10.1103/PhysRevLett.69.2863}.

\bibitem{Daley_2004}
A.~J. Daley, C.~Kollath, U.~Schollwöck and G.~Vidal,
\newblock \emph{Time-dependent density-matrix renormalization-group using adaptive effective hilbert spaces},
\newblock Journal of Statistical Mechanics: Theory and Experiment \textbf{2004}(04), P04005 (2004),
\newblock \doi{10.1088/1742-5468/2004/04/p04005}.

\bibitem{McCulloch_2007}
I.~P. McCulloch,
\newblock \emph{From density-matrix renormalization group to matrix product states},
\newblock Journal of Statistical Mechanics: Theory and Experiment \textbf{2007}(10), P10014 (2007),
\newblock \doi{10.1088/1742-5468/2007/10/P10014}.

\bibitem{to_mps3}
J.~I. Cirac, D.~Pérez-García, N.~Schuch and F.~Verstraete,
\newblock \emph{Matrix product states and projected entangled pair states: Concepts, symmetries, theorems},
\newblock Reviews of Modern Physics \textbf{93}(4) (2021),
\newblock \doi{10.1103/revmodphys.93.045003}.

\bibitem{to_mps4}
S.~R. White and A.~E. Feiguin,
\newblock \emph{Real-time evolution using the density matrix renormalization group},
\newblock Physical Review Letters \textbf{93}(7) (2004),
\newblock \doi{10.1103/physrevlett.93.076401}.

\bibitem{to_mps5}
U.~Schollwöck,
\newblock \emph{The density-matrix renormalization group in the age of matrix product states},
\newblock Annals of Physics \textbf{326}(1), 96–192 (2011),
\newblock \doi{10.1016/j.aop.2010.09.012}.

\bibitem{Kshetrimayum_Weimer_Orus_2017}
A.~Kshetrimayum, H.~Weimer and R.~Orus,
\newblock \emph{A simple tensor network algorithm for two-dimensional steady states},
\newblock Nature Communications \textbf{8}(1), 1291 (2017),
\newblock \doi{10.1038/s41467-017-01511-6},
\newblock ArXiv: 1612.00656.

\bibitem{Bañuls_Hastings_Verstraete_Cirac_2009}
M.~C. Bañuls, M.~B. Hastings, F.~Verstraete and J.~I. Cirac,
\newblock \emph{Matrix product states for dynamical simulation of infinite chains},
\newblock Physical Review Letters \textbf{102}(24), 240603 (2009),
\newblock \doi{10.1103/PhysRevLett.102.240603},
\newblock ArXiv: 0904.1926.

\bibitem{Czarnik_Dziarmaga_Corboz_2019}
P.~Czarnik, J.~Dziarmaga and P.~Corboz,
\newblock \emph{Time evolution of an infinite projected entangled pair state: an efficient algorithm},
\newblock Physical Review B \textbf{99}(3) (2019),
\newblock \doi{10.1103/PhysRevB.99.035115},
\newblock ArXiv: 1811.05497.

\bibitem{roy2025repulsivelyboundhadronsmathbbz2}
S.~G. Roy, V.~Sharma, K.~Xu, U.~Borla, J.~C. Halimeh and K.~R.~A. Hazzard,
\newblock \emph{{Repulsively Bound Hadrons in a ${Z}_{2}$ Lattice Gauge Theory}} (2025), \eprint{2510.23618}.

\bibitem{10.21468/SciPostPhys.16.5.138}
A.~Krasznai and G.~Takács,
\newblock \emph{{Escaping fronts in local quenches of a confining spin chain}},
\newblock SciPost Phys. \textbf{16}, 138 (2024),
\newblock \doi{10.21468/SciPostPhys.16.5.138}.

\bibitem{kondo-app}
J.~Chen, E.~M. Stoudenmire, Y.~Komijani and P.~Coleman,
\newblock \emph{Matrix product study of spin fractionalization in the one-dimensional kondo insulator},
\newblock Phys. Rev. Res. \textbf{6}, 023227 (2024),
\newblock \doi{10.1103/PhysRevResearch.6.023227}.

\bibitem{TDVP-1}
J.-W. Li, A.~Gleis and J.~von Delft,
\newblock \emph{Time-dependent variational principle with controlled bond expansion for matrix product states},
\newblock Phys. Rev. Lett. \textbf{133}, 026401 (2024),
\newblock \doi{10.1103/PhysRevLett.133.026401}.

\bibitem{URBANEK2016170}
M.~Urbanek and P.~Soldán,
\newblock \emph{Parallel implementation of the time-evolving block decimation algorithm for the bose–hubbard model},
\newblock Computer Physics Communications \textbf{199}, 170 (2016),
\newblock \doi{https://doi.org/10.1016/j.cpc.2015.10.016}.

\bibitem{Haegeman_Lubich_Oseledets_Vandereycken_Verstraete_2016}
J.~Haegeman, C.~Lubich, I.~Oseledets, B.~Vandereycken and F.~Verstraete,
\newblock \emph{Unifying time evolution and optimization with matrix product states},
\newblock Physical Review B \textbf{94}(16), 165116 (2016),
\newblock \doi{10.1103/PhysRevB.94.165116}.

\bibitem{Haegeman_Osborne_Verstraete_2013}
J.~Haegeman, T.~J. Osborne and F.~Verstraete,
\newblock \emph{Post-matrix product state methods: To tangent space and beyond},
\newblock Physical Review B \textbf{88}(7), 075133 (2013),
\newblock \doi{10.1103/PhysRevB.88.075133}.

\bibitem{PhysRevB.100.104303}
K.~H\'emery, F.~Pollmann and D.~J. Luitz,
\newblock \emph{Matrix product states approaches to operator spreading in ergodic quantum systems},
\newblock Phys. Rev. B \textbf{100}, 104303 (2019),
\newblock \doi{10.1103/PhysRevB.100.104303}.

\bibitem{PhysRevB.86.245107}
H.~N. Phien, G.~Vidal and I.~P. McCulloch,
\newblock \emph{Infinite boundary conditions for matrix product state calculations},
\newblock Phys. Rev. B \textbf{86}, 245107 (2012),
\newblock \doi{10.1103/PhysRevB.86.245107}.

\bibitem{Zauner_2015}
V.~Zauner, M.~Ganahl, H.~G. Evertz and T.~Nishino,
\newblock \emph{Time evolution within a comoving window: scaling of signal fronts and magnetization plateaus after a local quench in quantum spin chains},
\newblock Journal of Physics: Condensed Matter \textbf{27}(42), 425602 (2015),
\newblock \doi{10.1088/0953-8984/27/42/425602}.

\bibitem{Mi_2024}
X.~Mi \emph{et~al.},
\newblock \emph{Stable quantum-correlated many-body states through engineered dissipation},
\newblock Science \textbf{383}(6689), 1332–1337 (2024),
\newblock \doi{10.1126/science.adh9932}.

\bibitem{exp2}
C.~Nill, A.~Cabot, A.~Trautmann, C.~Gro\ss{} and I.~Lesanovsky,
\newblock \emph{Avalanche terahertz photon detection in a rydberg tweezer array},
\newblock Phys. Rev. Lett. \textbf{133}, 073603 (2024),
\newblock \doi{10.1103/PhysRevLett.133.073603}.

\bibitem{Wtwo_stepper}
M.~P. Zaletel, R.~S.~K. Mong, C.~Karrasch, J.~E. Moore and F.~Pollmann,
\newblock \emph{Time-evolving a matrix product state with long-ranged interactions},
\newblock Phys. Rev. B \textbf{91}, 165112 (2015),
\newblock \doi{10.1103/PhysRevB.91.165112}.

\bibitem{MPO-1}
E.~Mascarenhas, H.~Flayac and V.~Savona,
\newblock \emph{Matrix-product-operator approach to the nonequilibrium steady state of driven-dissipative quantum arrays},
\newblock Phys. Rev. A \textbf{92}, 022116 (2015),
\newblock \doi{10.1103/PhysRevA.92.022116}.

\bibitem{mpdo-2}
M.~V. Damme, J.~Haegeman, I.~McCulloch and L.~Vanderstraeten,
\newblock \emph{Efficient higher-order matrix product operators for time evolution} (2023), \eprint{2302.14181}.

\bibitem{mpdo-3}
F.~Verstraete, J.~J. Garc\'{\i}a-Ripoll and J.~I. Cirac,
\newblock \emph{Matrix product density operators: Simulation of finite-temperature and dissipative systems},
\newblock Phys. Rev. Lett. \textbf{93}, 207204 (2004),
\newblock \doi{10.1103/PhysRevLett.93.207204}.

\bibitem{Cui_Cirac_Bañuls_2015}
J.~Cui, J.~I. Cirac and M.~C. Bañuls,
\newblock \emph{Variational matrix product operators for the steady state of dissipative quantum systems},
\newblock Physical Review Letters \textbf{114}(22), 220601 (2015),
\newblock \doi{10.1103/PhysRevLett.114.220601},
\newblock ArXiv: 1501.06786.

\bibitem{mpdo-tebd}
H.~Weimer, A.~Kshetrimayum and R.~Or\'us,
\newblock \emph{Simulation methods for open quantum many-body systems},
\newblock Rev. Mod. Phys. \textbf{93}, 015008 (2021),
\newblock \doi{10.1103/RevModPhys.93.015008}.

\bibitem{mpdo-tebd2}
M.~Zwolak and G.~Vidal,
\newblock \emph{Mixed-state dynamics in one-dimensional quantum lattice systems: A time-dependent superoperator renormalization algorithm},
\newblock Phys. Rev. Lett. \textbf{93}, 207205 (2004),
\newblock \doi{10.1103/PhysRevLett.93.207205}.

\bibitem{complex-time}
J.~H. Cha, H.-Y. Lee and H.-S. Kim,
\newblock \emph{Dynamics of one-dimensional spin models via complex-time evolution of tensor networks},
\newblock Sci. Rep. \textbf{15}(1), 37490 (2025).

\bibitem{complex-time2}
M.~Grundner, P.~Westhoff, F.~B. Kugler, O.~Parcollet and U.~Schollw\"ock,
\newblock \emph{Complex time evolution in tensor networks and time-dependent green's functions},
\newblock Phys. Rev. B \textbf{109}, 155124 (2024),
\newblock \doi{10.1103/PhysRevB.109.155124}.

\bibitem{complex-time3}
X.~Cao, Y.~Lu, E.~M. Stoudenmire and O.~Parcollet,
\newblock \emph{Dynamical correlation functions from complex time evolution},
\newblock Phys. Rev. B \textbf{109}, 235110 (2024),
\newblock \doi{10.1103/PhysRevB.109.235110}.

\bibitem{complex-time-krylov}
S.~Paeckel,
\newblock \emph{Spectral decomposition and high-accuracy greens functions: Overcoming the nyquist-shannon limit via complex-time krylov expansion} (2026), \eprint{2411.09680}.

\bibitem{Childs_2021}
A.~M. Childs, Y.~Su, M.~C. Tran, N.~Wiebe and S.~Zhu,
\newblock \emph{Theory of trotter error with commutator scaling},
\newblock Physical Review X \textbf{11}(1) (2021),
\newblock \doi{10.1103/physrevx.11.011020}.

\bibitem{Hastings_2007}
M.~B. Hastings,
\newblock \emph{An area law for one-dimensional quantum systems},
\newblock Journal of Statistical Mechanics: Theory and Experiment \textbf{2007}(08), P08024–P08024 (2007),
\newblock \doi{10.1088/1742-5468/2007/08/p08024}.

\bibitem{brickwork1}
S.~Goto, R.~Kaneko and I.~Danshita,
\newblock \emph{Matrix product state approach for a quantum system at finite temperatures using random phases and trotter gates},
\newblock Phys. Rev. B \textbf{104}, 045133 (2021),
\newblock \doi{10.1103/PhysRevB.104.045133}.

\bibitem{brickwork2}
M.~DeCross, E.~Chertkov, M.~Kohagen and M.~Foss-Feig,
\newblock \emph{Qubit-reuse compilation with mid-circuit measurement and reset},
\newblock Phys. Rev. X \textbf{13}, 041057 (2023),
\newblock \doi{10.1103/PhysRevX.13.041057}.

\bibitem{Hydrodynamics1}
V.~Khemani, A.~Vishwanath and D.~A. Huse,
\newblock \emph{Operator spreading and the emergence of dissipative hydrodynamics under unitary evolution with conservation laws},
\newblock Phys. Rev. X \textbf{8}, 031057 (2018),
\newblock \doi{10.1103/PhysRevX.8.031057}.

\bibitem{Hydrodynamics2}
J.~Lux, J.~M\"uller, A.~Mitra and A.~Rosch,
\newblock \emph{Hydrodynamic long-time tails after a quantum quench},
\newblock Phys. Rev. A \textbf{89}, 053608 (2014),
\newblock \doi{10.1103/PhysRevA.89.053608}.

\bibitem{OTOCs}
C.~W. von Keyserlingk, T.~Rakovszky, F.~Pollmann and S.~L. Sondhi,
\newblock \emph{Operator hydrodynamics, otocs, and entanglement growth in systems without conservation laws},
\newblock Phys. Rev. X \textbf{8}, 021013 (2018),
\newblock \doi{10.1103/PhysRevX.8.021013}.

\bibitem{QGM1}
W.~S. Bakr, J.~I. Gillen, A.~Peng, S.~F{\"o}lling and M.~Greiner,
\newblock \emph{A quantum gas microscope for detecting single atoms in a hubbard-regime optical lattice},
\newblock Nature \textbf{462}(7269), 74 (2009).

\bibitem{topology_strings}
I.~Cong, N.~Maskara, M.~C. Tran, H.~Pichler, G.~Semeghini, S.~F. Yelin, S.~Choi and M.~D. Lukin,
\newblock \emph{Enhancing detection of topological order by local error correction},
\newblock Nat. Commun. \textbf{15}(1), 1527 (2024).

\bibitem{Jordan1928}
P.~Jordan and E.~Wigner,
\newblock \emph{{\"U}ber das paulische {\"a}quivalenzverbot},
\newblock Zeitschrift f{\"u}r Physik \textbf{47}(9), 631 (1928),
\newblock \doi{10.1007/BF01331938}.

\bibitem{sayak_guha_roy_2025_17479681}
S.~G. Roy,
\newblock \emph{Reweighted {TEBD} code repository},
\newblock \url{https://doi.org/10.5281/zenodo.18452815} (2025).

\end{thebibliography}

\nolinenumbers

\end{document}